\documentclass[preprint,showpacs,showkeys,preprintnumbers,amsmath,amssymb,aps]{revtex4}
\usepackage{epsfig}
\usepackage{graphicx}
\usepackage{rotating}
\usepackage{dcolumn}
\usepackage{bm}
\usepackage{tabularx}
\def\be{\begin{equation}}
\def\ee{\end{equation}}
\def\bea{\begin{eqnarray}}
\def\eea{\end{eqnarray}}
\def\barr{\begin{array}}
\def\earr{\end{array}}
\def\dis{\displaystyle}

\def\gev{\; {\rm GeV} }
\def\tev{\; {\rm TeV} }
\def\bea{\begin{eqnarray}}
\def\eea{\end{eqnarray}}

\def\bsub{\begin{subequations}}
\def\esub{\end{subequations}}

\def\f5g{{\cal F}_5^\gamma}
\def\f5z{{\cal F}_5^Z}
\def\h3g{{\cal H}_3^\gamma}
\def\h3z{{\cal H}_3^Z}
\def\tb{\tan{\beta}}
\begin{document}
\baselineskip 0.7cm
\title{New Physics Contribution to Neutral Trilinear Gauge Boson Couplings} 
\author{Sukanta Dutta$^{1}$}\email{Sukanta.Dutta@fnal.gov}
\author{ Ashok Goyal$^{2}$ }\email{agoyal@iucaa.ernet.in}
\author{Mamta$^{1}$ } \email{mdahiya@physics.du.ac.in}
\affiliation{$^1$SGTB Khalsa College, University of
Delhi. Delhi-110007. India.}
\affiliation{$^2$ Department of Physics \& Astrophysics, University of Delhi. Delhi-110007. India.}

\begin{abstract}
We study the one loop new physics effects to the CP even triple
neutral gauge boson vertices $\gamma^\star \gamma Z$, $\gamma^\star Z
\,Z$, $Z^\star \,Z\,\gamma$ and $Z^\star\,Z\,Z$ in the context of
Little Higgs models. We compute the contribution of the additional
fermions in Little Higgs models in the framework of direct product
groups where $[SU(2)\times U(1)]^2 $ gauge symmetry is embedded in
$SU(5)$ global symmetry and also in the framework of simple group
where $SU(N)\times U(1)$ gauge symmetry breaks down to $SU(2)_L\times
U(1)$. We calculate the contribution of the fermions to these
couplings when $T$ parity is invoked. In addition, we re-examine the
MSSM contribution at the chosen point of SPS1a' and compare with the
SM and Little Higgs models.
\end{abstract}

\maketitle
\section{Introduction}
\label{intro}
Multiple gauge boson production channels at the Large Hadron Collider
 (LHC) provide a novel opportunity to probe the trilinear and quartic
 gauge boson couplings \cite{Ball:2007zza}. The proposed International
 Linear Collider (ILC) will be much more sensitive to these couplings
 due to its clean environment and fixed center of mass energy
 \cite{Djouadi:2007ik}.  Availability of high luminosity in both these
 colliders gives us an unique facility to understand the non-Abelian
 gauge structure of the Standard Model (SM) and confront the presence
 of the new physics above the weak scale. The charged $WW\gamma$ and
 $WWZ$ couplings have been extensively studied and theoretical
 predictions in the context of SM and the minimal super-symmetric
 standard model (MSSM) have been made \cite{Gounaris:1996rz}. The
 neutral gauge-boson couplings $Z\gamma\gamma$, $ZZ\gamma$ and $ZZZ$
 which can be studied in $Z\gamma$ and $ZZ$ pair production in
 $e^+e^-$ and in hadron colliders through $e^+e^- \to Z \gamma,ZZ$ and
 $q \bar q \to Z\gamma, ZZ$ respectively have been analyzed within the
 SM and MSSM \cite{Gounaris:1999kf,Choudhury:2000bw}. Recently
 Armillis {\it et. al. } \cite{guzzi} performed a detailed study of
 the trilinear gauge boson interactions with additional anomalous
 $U(1)$'s, which arises in the construction of various string
 motivated and large extra dimension models. The model independent
 analysis of the neutral gauge boson couplings for hadron colliders
 exits in the literature \cite{baur} and has been also  recently
 studied with the Tevatron Data \cite{Deng:2008zz}. Recent LEP studies
 on the triple gauge boson couplings have been made in references
 \cite{:2007pq}.

\par The forthcoming experiments at the LHC and proposed ILC offer the
 exciting prospect of probing physics beyond SM. In particular if
 super-symmetry exists in nature, there will be a real possibility of
 discovering super-symmetric partners. As is well known super symmetry
 (SUSY) provides an elegant solution to the Hierarchy problem in the
 SM by canceling the quadratic divergence arising in the Higgs mass
 due to the contributions of SM particles by the contributions from
 their super partners.  

\par Recently there has been a proposal to
 consider Higgs fields as pseudo Nambu-Goldstone bosons of a global
 symmetry \cite{ArkaniHamed:2002qx,
 ArkaniHamed:2002qy,Low:2002ws,Skiba:2003yf,Kaplan:2003uc,Schmaltz:2004de}
 which is spontaneously broken at some high scale. The Higgs acquires
 mass through electro-weak symmetry breaking triggered by radiative
 corrections leading to a Coleman-Weinberg type of potential. Since
 the Higgs is protected by approximate global symmetry, it remains
 light and the quadratic contributions to its mass are canceled by the
 contributions of heavy gauge bosons and heavy fermionic states that
 are introduced in the model. If this little Higgs mechanism is
 realized in nature, it will be of great importance to verify it at
 the LHC. The realization of little Higgs mechanism discussed in the
 literature essentially fall into two classes
 \cite{Kaplan:2003uc}. The majority of implementations rely on the
 product group for the structure of gauge symmetry, its breaking
 pattern and the treatment of new heavy fermionic sector required to
 cancel the quadratic divergence contribution coming from the top
 quark. The second implementation has a simple group structure and has
 an additional model parameter. The Littlest Higgs model (LH)
 \cite{ArkaniHamed:2002qy} is a minimal model of the product group
 class which accomplishes this task to one loop order within a minimal
 matter content. SU(3) simple group model
 \cite{Kaplan:2003uc,Schmaltz:2004de} is a representative model of the
 second class. The SU(3) gauge symmetry in this model forces the
 introduction of heavy partners with each $SU(2)_L$ fermion doublets
 of the SM. This model has an extra parameter. Both these models
 however, suffer from severe constraints
 \cite{Csaki:2003si,Barbieri:2004qk,marandella} from precision
 electro-weak measurements which can be satisfied only by tuning the
 model parameters once again and thus introducing what is called a
 little hierarchy problem.

 Motivated by these considerations, an implementation of a discrete
 symmetry called T-parity is proposed. T-parity explicitly forbids any
 tree-level contribution from the heavy mass states to observables
 involving only the SM particles. It also forbids the interactions
 that impart vev to triplet Higgs, thereby generating the corrections
 to precision electro-weak observables only at the one loop level. In
 this little Higgs model with T-parity (LHT) \cite{Cheng:2003ju} there
 are heavy T-odd partners of the SM gauge bosons and SM fermions
 called mirror fermions. In the top quark sector, the model
 incorporates two heavy T-even and T-odd top quarks in addition to the
 T even SM top quark, which are required for canceling the
 quadratically divergent contribution of the SM top quark to the Higgs
 mass. The LHT has rich phenomenology and the LHC has great potential
 to unravel it by directly observing the T-partners of the SM
 particles as well as by studying indirect phenomenological
 consequences \cite{Hubisz:2004ft, Choudhury:2004bh}.

 In this paper we study the CP conserving trilinear neutral
 gauge-boson couplings in little Higgs models and MSSM as discussed above.
 In section II, we evaluate one loop fermion contribution to
 these three point functions when one of the gauge bosons is
 off-shell and in section III, we analyze and interpret the numerical results.


\section{Neutral gauge-boson couplings}
\label{ngbc}
 Bose-Einstein statistics render the three neutral gauge-boson
 couplings $\gamma\gamma Z$, $\gamma ZZ$ and $ZZZ$ to vanish when all
 the three vector bosons are on shell. The most general CP conserving
 coupling of one off-shell boson ${\rm V}\equiv Z/\, \gamma$ to a pair
 of on-shell $Z \gamma$ and $ZZ$ gauge bosons (all incoming) can be
 written as (see Ref.~\cite{Choudhury:2000bw})
\bsub
\begin{eqnarray}
\Gamma_{{\rm V} Z\gamma}^{\mu\alpha\beta}(Q,p_1,p_2)&=&i\,\,\left[{\cal H}_3^{\rm V}\epsilon^{\mu\beta\alpha\eta}\,\,{p_2}_\eta+
\frac{{\cal H}_4^{\rm V}}{M_Z^2}\,\,\Big\{\epsilon^{\mu\beta\rho\eta}\,\,{p_2}_{\rho}\,\,{Q}_\eta\,\,Q^{\alpha} \Big\} \right]\\
\Gamma_{{\rm V} ZZ}^{\mu\alpha\beta}(Q,p_1,p_2)&=&i\,\,\Big[{\cal F}_5^{\rm V}\epsilon^{\mu\alpha\beta\sigma}\,\,(p_1-p_2)_{\eta}\Big]
\end{eqnarray}
where the form factors ${\cal H}_i^{\rm V}$ and ${\cal F}_5^{\rm V}$
are related to those of reference \cite{hagiwara} by
\begin{eqnarray}
{\cal H}_i^{\gamma}=\frac{Q^2}{m_Z^2} h_i^{\gamma}, \hskip 0.2cm  
{\cal H}_i^{Z}=\frac{m_{Z}^2-Q^2}{m_Z^2} h_i^{Z} \hskip 0.3 cm {\rm and} \hskip 0.3 cm {\cal F}_5^{\rm V}=-\,\frac{Q^2-m_{\rm V}^2}{m^2_Z}\,\, f_5^V
\end{eqnarray}
\esub
Here $\Gamma_{{\rm V} V_1 V_2}^{\mu\alpha\beta}(Q,p_1,p_2)$ represents
the coupling of off-shell neutral gauge boson $V^\mu$ carrying
momentum $Q$ with the bosons $V_1^\alpha$ and $V_2^\beta$ carrying
momenta $p_1$ and $p_2$ respectively.

 In the SM these couplings vanish at the tree level. These
 couplings can however, be generated at the loop level Fig.~\ref{fig:feyn}).
 On account of the totally antisymmetric nature of
$\epsilon^{\mu\alpha\beta\sigma}$, these couplings can never
 be generated by scalars and vector-bosons running in the
 loop. Thus fermions running in the loop with one axial
 and two vector-couplings or all the three axial-couplings
 at the vertices can generate such couplings. Further at
 the one loop level, the couplings ${\cal H}_4^{\gamma}$ and
 ${\cal H}_4^Z$ are not generated i.e.
\begin{equation}
{\cal H}_4^{\gamma} = {\cal H}_4^Z = 0
\end{equation}
Thus the only couplings likely to appear at one-loop are ${\cal
 F}_5^{\gamma ,Z}$ and ${\cal H}_3^{\gamma,Z}$. These couplings can in
 general be complex quantities. However, they pick up imaginary
 contribution only when $Q^2$ crosses the threshold for fermion pair
 production (i.e. $Q^2 > 4m_f^2$) for time-like $Q^2$ or
 when $M_Z^2$ exceeds this threshold (i.e. $M_Z > 2m_f$) for space-like
 $Q^2$.

In order to evaluate these couplings, we write the
 interactions of the vector boson ${\rm V}\equiv\gamma ,\, Z$ with
 fermions in the standard notation
\begin{equation}
\mathcal L_{\rm int} =\bar f_i \,\, \gamma_{\mu}\,\,\Big[g^{\rm V}_{L_{ij}} \,\,P_L + g^{\rm V}_{R_{ij}}\,\,P_R\Big] f_j\,\, {\rm V}^{\mu}\, .
\end{equation}
For ${\rm V}\equiv \gamma$, we have $g_{L_{ij}}=
 g_{R_{ij}}=\delta_{ij}\,\,e\,\,q_i$, $q_i$ being the charge of
 fermion $f_i$. For ${\rm V}=Z$, the couplings $g_{_L}$ and $g_{_R}$
 in various models are listed in Table~\ref{tab:coup} of the Appendix. In
 the absence of any CP violating interactions, all these couplings are
 real and ${g_{L,R}}_{ij}={g_{L,R}}_{ji}$ because of hermiticity.

Using the notation of Passarino-Veltman (PV) functions, the
 contribution of the fermionic triangle graphs to the trilinear vector
 boson couplings ${\cal F}_5^{\gamma ,Z}$ and ${\cal H}_3^{\gamma,Z}$
 can be expressed in terms of scalar PV functions as given in the
 Appendix.

\subsection{SM contribution}

  The contribution to the trilinear neutral gauge couplings in the SM
 arise from the three families of quarks and leptons. The anomaly
 cancellation ensures that all the couplings go to zero for $Q^2$ much
 larger than the fermion pair production threshold. It is obvious that
 of all the thresholds (at $Q^2=4\,\,M_f^2$), the largest contribution
 comes from the heaviest fermion loop. The couplings ${\cal
 F}_5^{\gamma }$ and ${\cal H}_3^{\gamma,Z}$ get contributions only
 from the charged fermions, whereas ${\cal F}_5^Z$ receives
 contributions from the neutrinos as well. Note that in SM, the same
 fermion runs in all the three sides of triangle loop as there is no
 mixed coupling of neutral boson with different fermions.

\subsection{MSSM}
The MSSM contribution to the trilinear neutral gauge couplings has
been calculated in the references \cite{Choudhury:2000bw,
Gounaris:1999kf}.  Charginos contribute to all the four anomalous
couplings where as the neutralinos contribute only to the ${\cal
F}_5^Z$.  We re-calculate the MSSM contribution in the light of the
reference point SPS1a' which is defined at a characteristic scale of 1
TeV with its origin in minimal super-gravity (mSUGRA) \cite{spa}. The
root GUT scale mSUGRA parameters in this reference point SPS1a' are
the gaugino mass $M_{1/2}=250$ GeV, the universal scalar mass $M_0=70
$ GeV, the trilinear coupling $A_0=-300$ GeV, $\tan\beta({\tilde
M})=10$ and sign($\mu)=+1$. Extrapolating these parameters to
$\tilde{M}=1$ TeV generates the MSSM Lagrangian parameters. The
relevant evolved MSSM parameters for our calculations are the Higgs
mixing parameter $\mu=396$ GeV and $M_2=193.2$ GeV.
\subsection{ LH contribution}
 In the little Higgs models, the Higgs bosons are realized as pseudo-Goldstone
bosons. The generic structure of little Higgs models is a global
symmetry broken at a high (TeV) scale $f$. At this scale there are new
gauge bosons, scalars and fermions responsible for the  cancellation of the
quadratic divergent one loop contributions to the Higgs boson mass
from the SM gauge bosons, Higgs self interactions and from the top
quark respectively. The Littlest Higgs model accomplishes this task
with the minimal matter content. In this model $[SU(2)\times U(1)]^2$
gauge symmetry is embedded in an $SU(5)$ global symmetry. The gauge
symmetry is broken down to the SM $SU(2)\times U(1)$ gauge symmetry by
a single vacuum condensate $f\approx $ 1 TeV. The new fermionic degrees of
freedom in the Littlest Higgs model are in the heavy quark sector and
consist of a pair of vector-like $SU(2)$-singlet quarks that couple to
the top sector. The resultant top sector consists of a top quark $t$
and its heavy partner T whose masses and couplings are given in terms
of model dependent  parameters by
\begin{eqnarray}
m_t &=& \frac{\lambda_1\,\, \lambda_2}{\sqrt{\lambda_1^2+\lambda_2^2}}\,\, v   \\
M_T &=& \sqrt{\lambda_1^2+\lambda_2^2}\,\,
f = \frac{1}{\sqrt{X_L(1-X_L)}}\,\,\frac{m_t}{v}\,\, f
\end{eqnarray}
where $X_L=\lambda_1^2/\,\big(\lambda_1^2+\lambda_2^2\big)$,
$\lambda_1$ and $\lambda_2$ being the couplings that appear in the
heavy quark sector of the interaction lagrangian.
%
\begin{table}[!h]
\begin{center}
\begin{tabular}{||c|c|c||c|c|c|c||}
\hline
\hline
&&&&&&\\[-3mm]
scale & $m_T$ & $\sqrt{Q^2}$ & ${\cal H}_3^\gamma$ & ${\cal H}_3^Z$ & ${\cal F}_5^\gamma$ 
& ${\cal F}_5^Z$ \\
(in TeV)&(in GeV)& &$(10^{-4})$ &$(10^{-4})$&$(10^{-4})$&$(10^{-4})$ \\
\hline\hline
 & $ $ &  $2m_t $ &  $-89.34 - \iota \,0.0130$ &  $ 25.49 + 0\,\iota $ 
&  $-27.65 + 0\,\iota $ &  $ -18.06 + 0\, \iota \,$ \\
$0.5$&  $711.4 $ &  $m_t+M_{T} $ &  $ 0.9154 -\iota \,28.02 $ &  
$ 5.388 + \iota \,7.730$ &  $ -1.313 - \iota \,9.229$ &  
$ -8.534 - \iota \,11.50$ \\
&  $ $ &  $2M_{T} $ &  $3.710 - \iota \,14.24 $ &  $ -0.3113 +\iota \,6.901$ 
&  $ -0.0809 - \iota \,5.481$ &  $ 1.717 -\iota \,10.66$ \\
\hline
 & $ $ &  $2m_t $ &  $ -92.97 -\iota \,0.0147$ &  $ 28.46 + \iota 0.0003$ 
&  $ -30.40 +0\,\iota $ &  $ -19.98 +\iota \,0.0002$ \\
$1.0$&  $ 1422.8 $ &  $m_t+M_{T} $ &  $ 4.6901 -\iota \,12.51 $ &  
$ -0.1816 +\iota \,3.951$ &  $  1.152 - \iota \,4.267$ &  
$ -0.5294- \iota \,6.851$ \\
&  $ $ &  $2M_{T} $ &  $ 2.976 - \iota \, 5.004$ &  $ -0.842 +\iota \,2.255$ 
&  $  0.6727 - \iota \,1.942 $ &  $ 1.710 -\iota \,3.926$ \\
\hline
 & $ $ &  $2m_t $ &  $ -93.87 -\iota \,0.0152$ &  $29.26  + \iota \, 0.0008$ 
&  $ -31.22 + 0\, \iota $ &  $ -20.57 +\iota \,0.0002$ \\
$2.0$&  $2845.5 $ &  $m_t+M_{T} $ &  $  3.039 - -\iota \,4.590 $ &  
$ -0.7061 + \iota \,1.501$ &  $ 0.9257 - \iota \,1.573$ &  
$  0.9056- \iota \,2.891$ \\
&  $ $ &  $2M_{T} $ &  $ 1.396 -\iota \,1.587 $ &  $ -0.4402 +\iota \,0.6802$
&  $ 0.3927 - \iota \,0.6076$ &  $ 0.8645 - \iota \,1.2824$ \\
\hline
 & $ $ &  $2m_t $ &  $-94.04 - \iota \,0.0152$ &  $ 29.41 +0\, \iota $ 
&  $ -31.37 +0\, \iota $ &  $ -20.68 +0 \,\iota \,$ \\
$3.0$&  $4268.3 $ &  $m_t+M_{T} $ &  $  0.8188 -\iota \,0.7906 $ &  
$0.2639 +  \iota \,0.3314$ &  $ 0.2422 - \iota \,0.2998$ &  
$  0.5222  \iota \,0.6467$ \\
&  $ $ &  $2M_{T} $ &  $ 1.9801 - \iota \, 2.4200$ &$-0.5342 +\iota \,0.7975$
&  $  0.6259 -\iota \, 0.8290$ &  $ 0.7985 -\iota \,1.6052$ \\
\hline
\hline
\end{tabular}
\caption{ \label{tab:lht_scale_var} {\em The values of various triple
neutral gauge boson couplings in LH Model (written as complex numbers)
at some typical $\sqrt{Q^2}$ (where peaks are expected) for different
values of symmetry breaking scale $f$.  All values correspond to $r =
\lambda_1/ \lambda_2= 1$ and $m_t = 175$GeV. The values are written in
units of $10^{-4}$. Note that the lower values of the scale correspond
only to the Littlest Higgs Model with T-Parity (the couplings in the
two models being the same up to ${\cal O}(v/f)$) with $m_{T_+}= m_T$.}}
\end{center}
\end{table}
\par The interactions of the left and right handed fermions with the
$Z$ boson in this model can be found in the reference
\cite{Han:2005ru} and can also be realized from the LHT couplings given in
the Table \ref{tab:coup} by retaining only the leading terms in
$v/f $ because of the large scale factor requirement from
precision electro-weak data.
\par The second class of little Higgs models feature a simple
 group that contains an $SU(N)\times U(1)$ gauge symmetry that is
 broken down to $SU(2)_L\times U(1)$, giving rise to a set of
 TeV-scale gauge bosons. The two gauge couplings of $SU(N)\times U(1)$
 are fixed in terms of two SM gauge couplings, leaving no free
 parameters in the gauge sector. Furthermore, due to enlarged $SU(N)$
 gauge symmetry, all fermionic SM representations are extended to
 transform as fundamental or conjugate representations of
 $SU(N)$. This gives rise to additional heavy fermions in all the
 three quark and lepton sectors. The simplest realization of this
 simple group class is the $SU(3)$ simple gauge model
 \cite{Kaplan:2003uc,Schmaltz:2004de} with anomaly-free embedding of
 extra fermions. The expansion of the $SU(2)_L$ gauge group to $SU(3)$
 requires introduction of heavy fermions associated with each
 $SU(2)_L$ doublet of the SM. The first two generations of quarks are
 enlarged to contain the new TeV-scale D and S quarks of charge $-1/3$
 that are $\bar 3$ representations of $SU(3)$. The quarks of
 the third generation and three generation of leptons are put in the 3
 representation of $SU(3)$. The electric charge of the heavy third
 generation quark T is $+2/3$ and all the heavy leptons $N_i$ of three
 generations are electrically neutral. The masses of these heavy
 fermions are given in terms the parameters of the model, namely,
\begin{eqnarray}
M_T &=& \sqrt{\lambda_1^2\,\, c_{\beta}^2 + \lambda_2^2\,\,s_{\beta}^2}\,\, f = \sqrt{2}\,\, \frac{t_{\beta}^2+X_{\lambda}^2} 
{(1+t_{\beta}^2)\,\, X_{\lambda}}\,\, \frac{m_t}{v}\,\,f  \nonumber \\
M_{D,\,S} &=& s_{\beta}\,\,\lambda_{D,\,S}\,\, f \nonumber \\
M_{N_i} &=& s_{\beta}\,\,\lambda_{N_i}\,\,f 
\end{eqnarray}
 where $t_{\beta} \equiv \tan\beta=f_2/\, f_1$ is an additional
parameter in the simple $SU(3)$ model and $X_L=\lambda_1^2/\,
\left(\lambda_1^2+ \lambda_2^2\right)$. in $SU(3)$ simple group, $f$
is defined as $f \equiv \sqrt{f_1^2 + f_2^2}$. In these expressions
effect of light quark masses is neglected and the neutrinos are taken
to be mass less.  Constraints from electro-weak precision measurements
require the breaking scale $f$ to be greater than 5 TeV in the
Littlest Higgs Model. This constraint can however be brought down to
about $2-3$~TeV ( See for example Ref. \cite{Han:2005ru} ). In the
anomaly free $SU(3)$-simple group~\cite{marandella}, the constraint on
the scale is $f>$ 3.9 TeV for $t_{\beta}$ = 3. The scale $f$ can only
be marginally brought down by slightly different realization
\cite{simplemod1}. In these two classes of models $T$ mass has a lower
bound given by
\begin{eqnarray}
M_T \geq 2\,\,\frac{m_t}{v}\,\,f \approx \sqrt{2}\,\, f\,\,\,\, {\rm for}\,\, \lambda_1=\lambda_2
\end{eqnarray}
in the Littlest Higgs model and
\begin{eqnarray}
M_T \geq 2\,\,\sqrt{2} \,\, s_{\beta}\,\,c_{\beta}\,\,\frac{m_t}{v}\,\, f\approx f\,\,\sin (2\,\beta )  \,\,\,\, {\rm for}\,\, \frac{\lambda_1}{\lambda_2}=\tan\beta
\end{eqnarray}
 in the $SU(3)$ simple group model.  
\begin{table}[!h]
\begin{center}
\begin{tabular}{||c||c|c|c|c||}
\hline
\hline
&&&&\\[-3mm]
$\sqrt{Q^2}$ & ${\cal H}_3^\gamma$ & ${\cal H}_3^Z$ & ${\cal F}_5^\gamma$ 
& ${\cal F}_5^Z$ \\
(in TeV)&$(10^{-4})$ &$(10^{-4})$&$(10^{-4})$&$(10^{-4})$ \\
\hline\hline
$2m_t  $ &   $-94.17 -\iota \, 0.0158 $ &  $29.53 + 0 \iota\, $ &
$-31.50+\iota \,0.0149$ &  $ -22,.42 +\iota \, 0.0254 $ \\
\hline
$m_t+M_T$ &  $4.533 -\iota \, 9.136$ &  $-1.487 + \iota \, 3.008$ &  
$ 1.757- \iota\, 2.802$ & $ -0.1062 -\iota \, 4.751$ \\
\hline
$2 M_T$ &   $2.582- \iota\,  3.448 $ &  $ -2.417 -\iota\, 0.0712 $ &  
$ 0.5535- \iota\, 0.677$ &  $ 0.9483 +\iota \, 0.9624$ \\
\hline
$M_U$ &   $3.146 - \iota \,4.660 $ &  $ -2.503 +\iota \, 2.036$ &  
$1.146 -\iota \,1.152$ &  $ 1.699 - \iota \, 2.449$ \\
\hline
$2M_U$ &  $1.372 -\iota \,1.455$ &  $ 0.1424 - \iota \,1.523$ &  
$ 2.947- \iota \,0.191 $ &  $ -3.151 +\iota \, 2.236$ \\
\hline
\hline
\end{tabular}
\caption{ \label{tab:su3} {\em The values of various couplings
    (written as complex numbers) at some typical $\sqrt{Q^2}$ (where
    peaks are expected)  in the SU(3) simple model with anomaly free
    embedding. All values correspond to $\tan{\beta} = r = 3$, scale
    $f = 3$TeV and $m_t = 175$GeV. At these values of parameters, the
    mass of heavy top is $M_T = 1.8$~TeV and masses of all other heavy
    fermions have been taken to be $M_i= 3$~TeV.}}
\end{center}
\end{table}
\begin{figure}
\begin{center}
\hskip -3 cm
  \begin{minipage}[t]{0.33\textwidth}
   \includegraphics[width=9 cm,height=14 cm,]
   {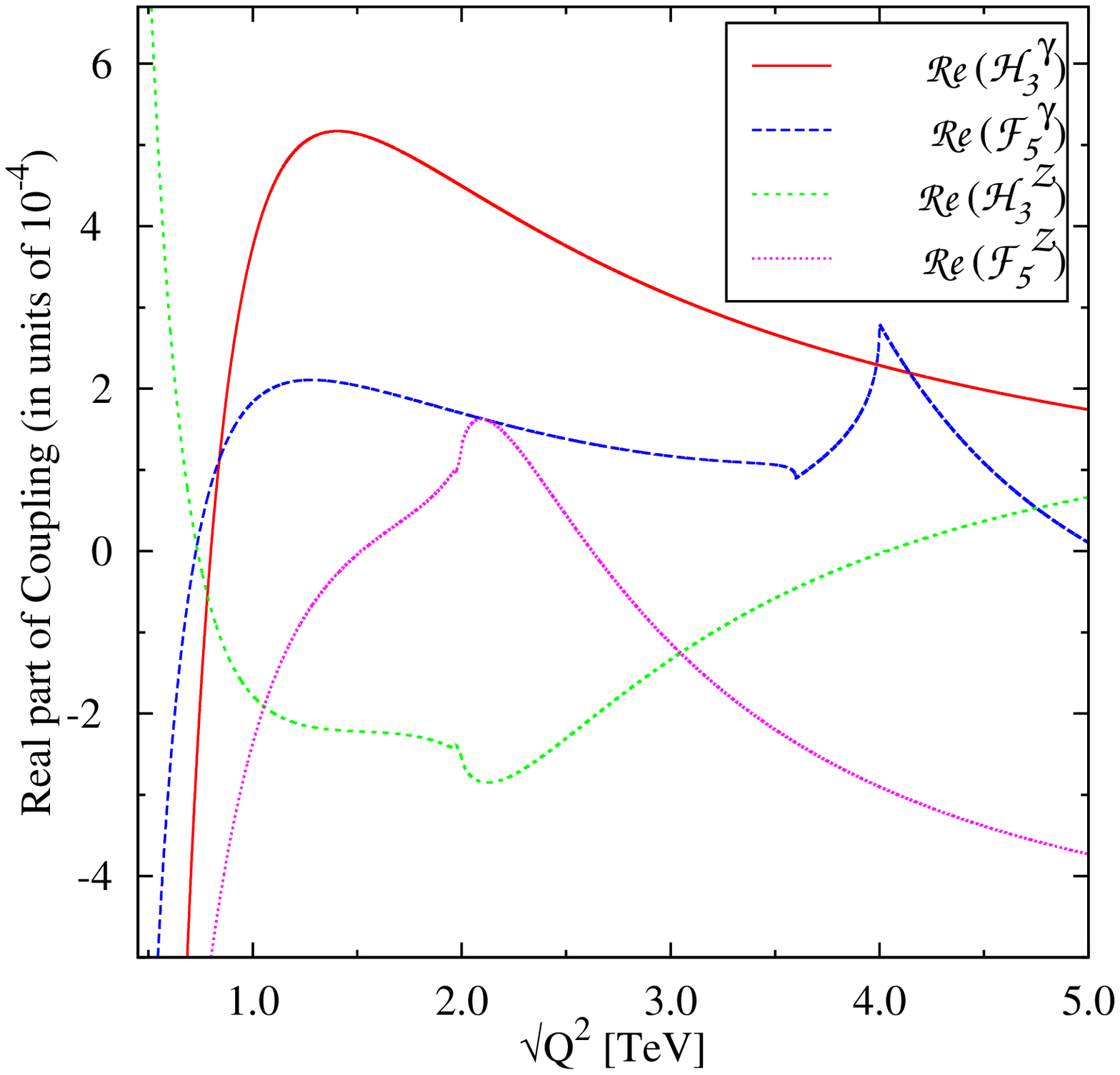}
  \end{minipage}
\hskip 3 cm
  \begin{minipage}[t]{0.33\textwidth}
   \includegraphics[width=9cm,height=14 cm]
   {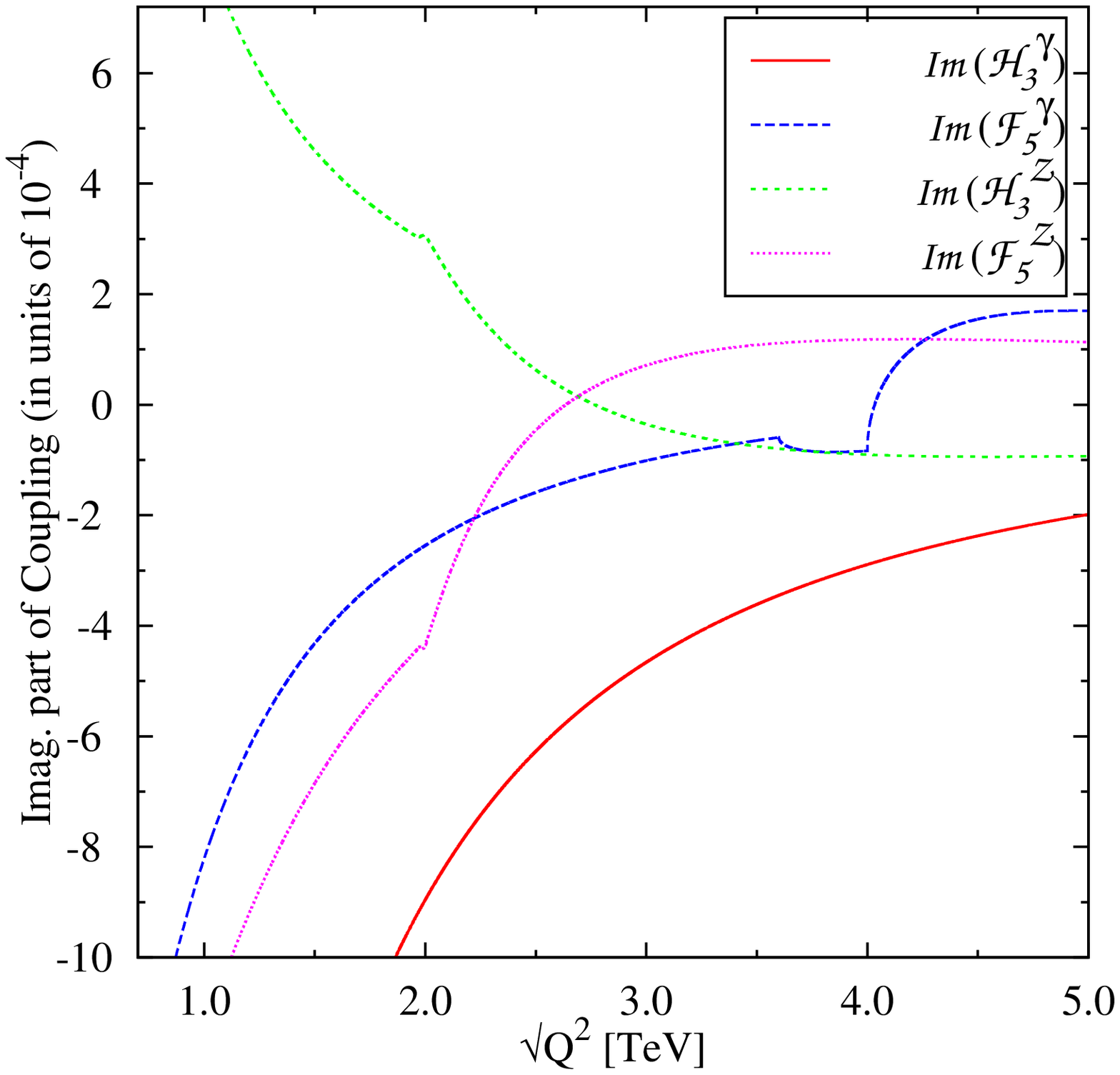}
   \end{minipage}
\vskip -3cm
\caption{\small\em $\sqrt{Q^2}$-variation in the range $1-5 \tev$ of the 1)
real and 2) imaginary parts of the couplings in SU(3) simple Little
Higgs Model with anomaly free embedding for $\tb= r = 3$, $f=3
\tev$. With this choice of parameters the mass of heavy top $M_T = 1.8
\tev$ while the masses of all other heavy fermions are fixed at $2
\tev$.}
 \label{su3}
\end{center}
\end{figure}

\par We calculate the contributions to the trilinear gauge boson
couplings in the LH model. Unlike SM, here the contribution to these
couplings also comes from the triangle graph with both the SM top $t$
and heavy top $T$ simultaneously present in the loop because of the
presence of $\bar T Zt$ coupling.  The couplings relevant for our study
can be read out from the Table~\ref{tab:coup} of the Appendix.

In Table~\ref{tab:lht_scale_var}, we give the values of real and
imaginary parts of all four trilinear neutral gauge couplings as a
function of symmetry breaking scale $f$ for fixed ratio
$r=\frac{\lambda_1}{\lambda_2} = 1$. The values are given for some
typical values of $\sqrt{Q^2}$ where theoretical peaks are expected.
It is worthwhile to mention once again, that the entries corresponding to the large values for
the scale $f$ say, $\sim 2-3$~TeV  corresponds to both  the little Higgs Model with and without T Parity. However, the results derived from  the lower values of the scale $f$ corresponds only to the  Little Higgs Model with T parity (discussed later in the following section) as the measured value  of precision observables forbids lower value of $f$ in the littlest Higgs model. It is to be noted  that up to the leading order in $x={\cal O}(v/f)$, the couplings in both the
models are same and for large $f$ the higher order terms in  $x$ may be easily
neglected.

We also calculate the contributions to the trilinear gauge boson
couplings in the anomaly free SU(3) simple group model. A new feature
in this model is the contribution from mixed $t$ and $T$, mixed SM and
TeV range quarks of the first two generations and the mixed neutrino
and TeV mass heavy neutrinos $(N_i)$ of all the three generations in
the triangle loop. However, the pure $T$ quark loop, pure TeV mass
quark loop of first two generations and TeV mass heavy neutrinos of
three generations do not contribute to ${\cal F}_5^Z$ in the
model. This is clear from the couplings of Z-boson to various new
fermions (Table~\ref{tab:coup}).

The Table~\ref{tab:su3} lists the values of the couplings for the
anomaly free SU(3) simple group Model. All values correspond to
$\tan{\beta} = r = 3$, scale $f = 3$~TeV and $m_t = 175$~GeV. At these
values of the parameters, the mass of heavy top is $M_T = 1.8$~TeV and
masses of all other heavy fermions have been taken to be $M_i= 3$~TeV.
As expected, the threshold values at $2m_t$ has roughly the same
magnitude as that in the SM. At higher $\sqrt{Q^2}$, the effect of new
heavy fermions shows up but the threshold values are an order of
magnitude lower than that at the $2m_t$ threshold. However, at these
$\sqrt{Q^2}$, the SM contribution is negligible.
 In Fig.~\ref{su3} we have shown this behavior of couplings as a
function of $\sqrt{Q^2}$. The values of various parameters are the
same as given in Table~\ref{tab:su3} but the mass of heavy fermions
$U$ and $N$ which are taken to be $2\tev$ each in the figure.
\subsection{LHT contribution}
In LHT as discussed in the Introduction, the $T$-odd heavy (TeV mass)
 fermions called mirror fermions couple vectorially to
 $Z_{\mu}$. Further, because of $T$-parity conservation there is no
 coupling between $Z_{\mu}$ and T-odd and $T$-even fermions i.e. the
 coupling $Z\bar f_+f_- =0$ Thus the mirror fermions do not contribute
 to the trilinear neutral gauge boson couplings.  The $T$-even partner
 of the top quark $T_+$ however, has both axial and vector couplings
 with $Z_{\mu}$ and hence contributes to the triangle loop.  The top
 quark masses in this model are given by \bsub
\begin{eqnarray}
m_t &=& \frac{\lambda_1\lambda_2}{\sqrt{\lambda_1^2+\lambda_2^2}}\,\,v\,\,\left\{ 1+\frac{v^2}{f^2}\,\left( -\frac{1}{3} +\frac{1}{2}\, X_L\,\, (1-X_L)  \right)\right\}  \\
M_T &=& \frac{m_t}{\sqrt{X_L\,(1-X_L)}}\,\,\frac{f}{v}\,\, \left\{ 1+\frac{v^2}{f^2}\,\left( \frac{1}{3} - X_L\,\, (1-X_L)  \right)\right\} 
\end{eqnarray}
\esub
\begin{figure}[!h]
\vskip -2cm
\begin{center}
   \includegraphics[width=10cm,height=14 cm]{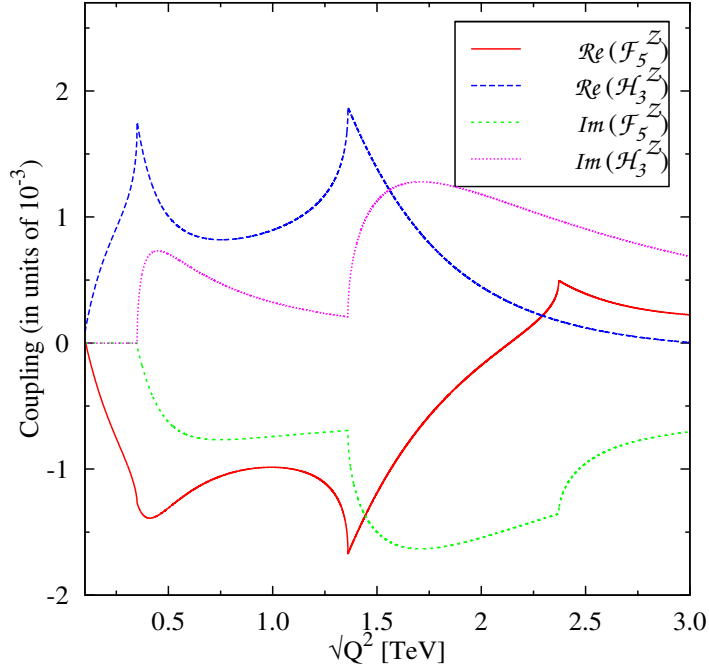}
\vskip -3cm
\caption{\small\em $\sqrt{Q^2}$-variation of real and imaginary parts of
$\h3z$ and $\f5z$ in the range $0-3 \tev$  in Little Higgs Model with T-parity
 for $ r = 3$ and $f=0.5 \tev$. With this
choice of parameters the mass of T-even top, $M_{T_+} = 1.186 \tev$.}
  \label{lht_rat3}
\end{center}
\end{figure}
\par
 In addition to the contribution from SM fermions in the loop, we now
 get additional contributions from heavy $T$-even partner and the top quark as
 well as contributions from the triangle loops with mixed
 contributions from $t$ and $T+$ quarks. The relevant fermion couplings
 with $Z$ boson are given in table \ref{tab:coup} of the appendix.
\begin{table}[!h]
\begin{center}
\begin{tabular}{||c|c|c||c|c|c|c||}
\hline \hline &&&&&&\\[-3mm] ratio & $M_{T_+}$ & $\sqrt{Q^2}$ &
${\cal H}_3^\gamma$ & ${\cal H}_3^Z$ & ${\cal F}_5^\gamma$ & ${\cal F}_5^Z$ \\
 &(in GeV)&&$(10^{-4})$ &$(10^{-4})$&$(10^{-4})$&$(10^{-4})$ \\ 
\hline\hline & $
$ & $2m_t $ & $-93.40 - \iota \, 0.0149 $ & $ 28.85 + 0 \,\iota \,$ &
$-30.83 +0 \,\iota $ & $ -20.29 + 0 \,\iota \,$ \\ $0.5$& $ 889.2 $ &
$m_t+M_{T_+} $ & $ 4.068 -\iota \, 22.963$ & $ -0.705 + \iota \,
7.341$ & $1.413 -\iota \,7.612 $ & $ -1.601 - \iota \,10.82$ \\ & $ $
& $2M_{T_+} $ & $4.595 -\iota \, 10.66 $ & $ -1.499 + \iota \, 3.875 $
& $ 1.441- \iota \,3.666$ & $ 1.574 -\iota \,6.475$ \\ 
\hline & $ $ &
$2m_t $ & $-89.34 - \iota \,0.0130$ & $ 25.49 + 0\,\iota $ & $-27.65 +
0\,\iota $ & $ -18.06 + 0\, \iota \,$ \\ $1.0$& $711.4 $ &
$m_t+M_{T_+} $ & $ 0.9154 -\iota \,28.02 $ & $ 5.388 + \iota \,7.730$
& $ -1.313 - \iota \,9.229$ & $ -8.534 - \iota \,11.50$ \\ & $ $ &
$2M_{T_+} $ & $3.710 - \iota \,14.24 $ & $ -0.3113 +\iota \,6.901$ & $
-0.0809 - \iota \,5.481$ & $ 1.717 -\iota \,10.66$ \\ 
\hline & $ $ &
$2m_t $ & $ -81.80 - \iota \,0.0094$ & $ 19.84 - \iota \,0.0054$ & $
22.00 + 0\, \iota \,$ & $-14.31 + \iota \,0.0002 $ \\ $2.0$& $889.2 $
& $m_t+M_{T_+} $ & $0.7583 - \iota \,19.59 $ & $14.88 + \iota \,3.813
$ & $-4.727 - \iota \,7.226 $ & $ -15.09 - \iota \,8.566$ \\ & $ $ &
$2M_{T_+} $ & $1.116 - \iota \,9.099 $ & $1.424 + \iota \,9.069$ &
$-2.866 - \iota \,5.549 $ & $ -5.549 - \iota \,13.21$ \\ 
\hline & $ $
& $2m_t $ & $ -78.51 - \iota \,0.0079$ & $17.58 -\iota \,0.0068 $ & $
-19.40 + 0 \,\iota $ & $-12.69 + \iota \, 0.0002$ \\ $3.0$& $1185.6 $
& $m_t+M_{T_+} $ & $0.4505 - \iota \,13.03$ & $18.70 +\iota \,2.076 $
& $-6.278 - \iota \,6.100 $ & $ -16.71 - \iota \,6.937$ \\ & $ $ &
$2M_{T_+} $ & $-0.623 - \iota \,5.49$ & $1.759 +\iota \, 9.726$ & $
-3.972 -\iota \,5.508$ & $ 4.953 - \iota \,13.55$\\ 
\hline & $ $ &
$2m_t $ & $-77.05 - \iota \,0.0072$ & $ 16.60 -\iota \,0.0074$ & $
-18.16 + 0\,\iota \,$ & $-11.95 + \iota \,0.0002$ \\ $4.0$& $1511.7 $
& $m_t+M_{T_+} $ & $ 0.2019 - \iota \,9.182$ & $ 20.18 +\iota \,1.320$
& $ -7.086- \iota \,5.541$ & $ -17.0 - \iota \,6.011$ \\ & $ $ &
$2M_{T_+} $ & $-1.712 - \iota \,3.614$ & $ 1.824 + \iota \,9.939$ &
$-4.461 - \iota \,5.508 $ & $ 5.479 - \iota \,13.48$ \\ 
\hline & $ $ &
$2m_t $ & $ -76.30 - \iota \,-0.0068$ & $ 16.10 - \iota \,0.0077$ &
$-17.49 +0 \,\iota \, $ & $ -11.56 + \iota \,0.0002$ \\ $5.0$& $
1849.6 $ & $m_t+M_{T_+} $ & $-0.843 - \iota \,6.819$ & $20.78 + \iota
\, 0.9261$ & $ -7.562 - \iota \,5.220$ & $-17.14 - \iota \,5.422 $ \\
& $ $ & $2M_{T_+} $ & $-2.416 - \iota \,2.558$ & $1.841 + \iota \,
10.01$ & $ -4.717 - \iota \,5.511$ & $ 5.727 - \iota \,13.34$ \\
\hline \hline
\end{tabular}
\caption{ \label{tab:lht_rat_var} {\em The values of various couplings
(written as complex numbers) at some typical $\sqrt{Q^2}$ (where peaks
are expected) for different values of $r = \lambda_1/ \lambda_2$ in
the LHT Model. All values correspond to symmetry breaking scale $f=
500$GeV and $m_t = 175$GeV.}}
\end{center}
\end{table}

The Table~\ref{tab:lht_rat_var} lists the values of the couplings in
the Little Higgs Model for different values of the $r$ ratio and at
some typical values of $\sqrt{Q^2}$. All the values are for symmetry
breaking scale $f=500 \gev$ and are given in units of $10^{-4}$. This
table also gives the value of heavy T-even top mass for different
ratios. As expected for the same value of $f$, the mass $m_{T_+}$ is
the same for the ratio $r$ and $1/r$. It may be mentioned that for
higher ratios, a very interesting behavior is shown by the couplings
$\h3z$ and $\f5z$. Not only the imaginary part becomes appreciable at
high $\sqrt{Q^2}$ but also the threshold values of the couplings at
$\sqrt{Q^2} = m_t + m_{T_+}$ are higher than those at $\sqrt{Q^2} = 2
m_t$ and are comparable to the SM values. This is clearly brought
about in Fig. \ref{lht_rat3} where we have plotted the real and
imaginary parts of the couplings ${\cal H}_3^Z$ and ${\cal F}_5^Z$ for
$\sqrt{Q^2}$ up to 3 TeV, scale $f$ = 0.5 TeV and the ratio $r$ = 3.


\section{Results and Discussion}
\label{results}

\begin{figure}[tbh!]
 \begin{center}
\hskip -3 cm
  \begin{minipage}[t]{0.33\textwidth}
   \includegraphics[width=8 cm,height=12 cm,]
   {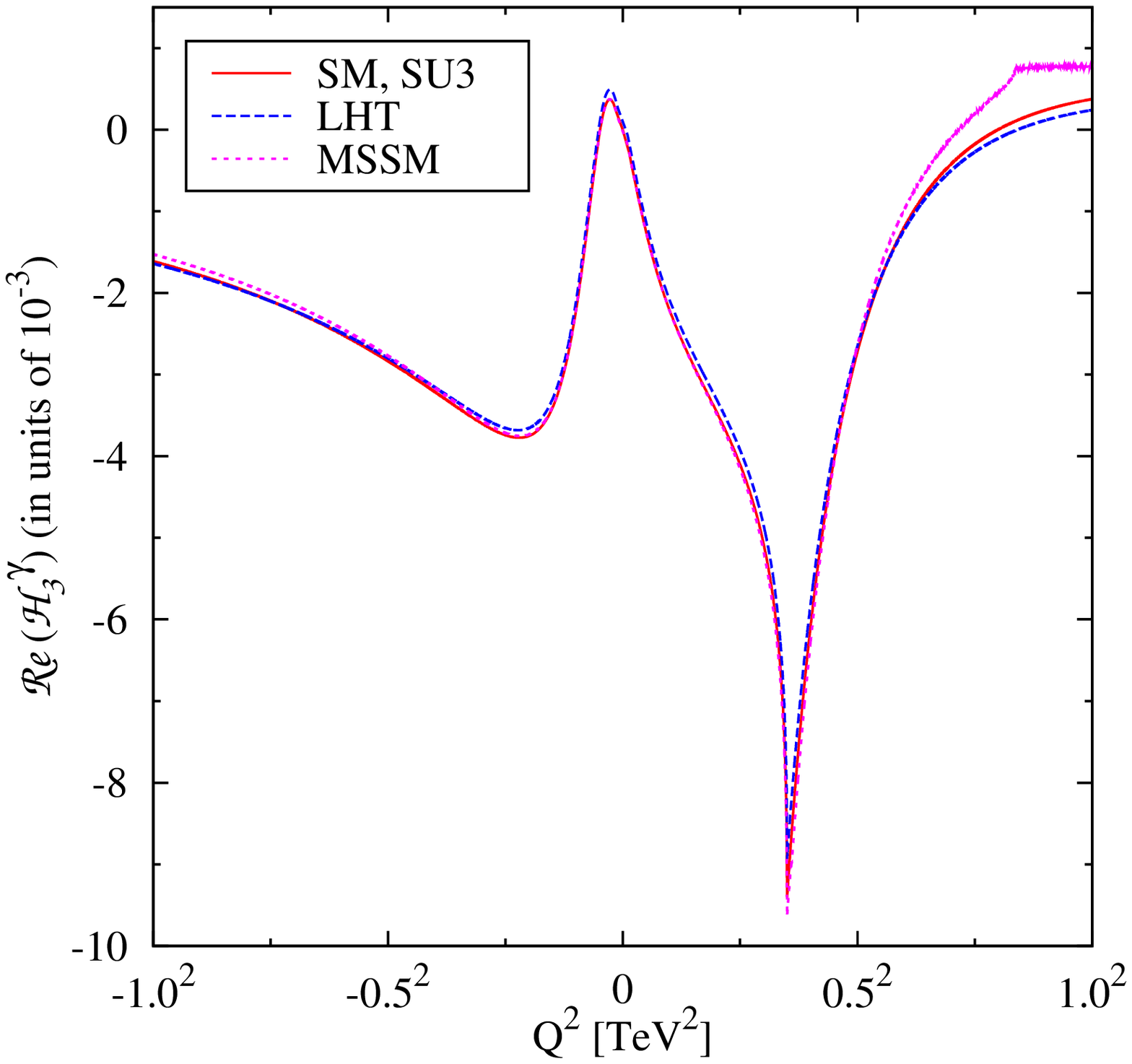}
  \end{minipage}
\hskip 3 cm
  \begin{minipage}[t]{0.33\textwidth}
   \includegraphics[width=8cm,height=12 cm]
   {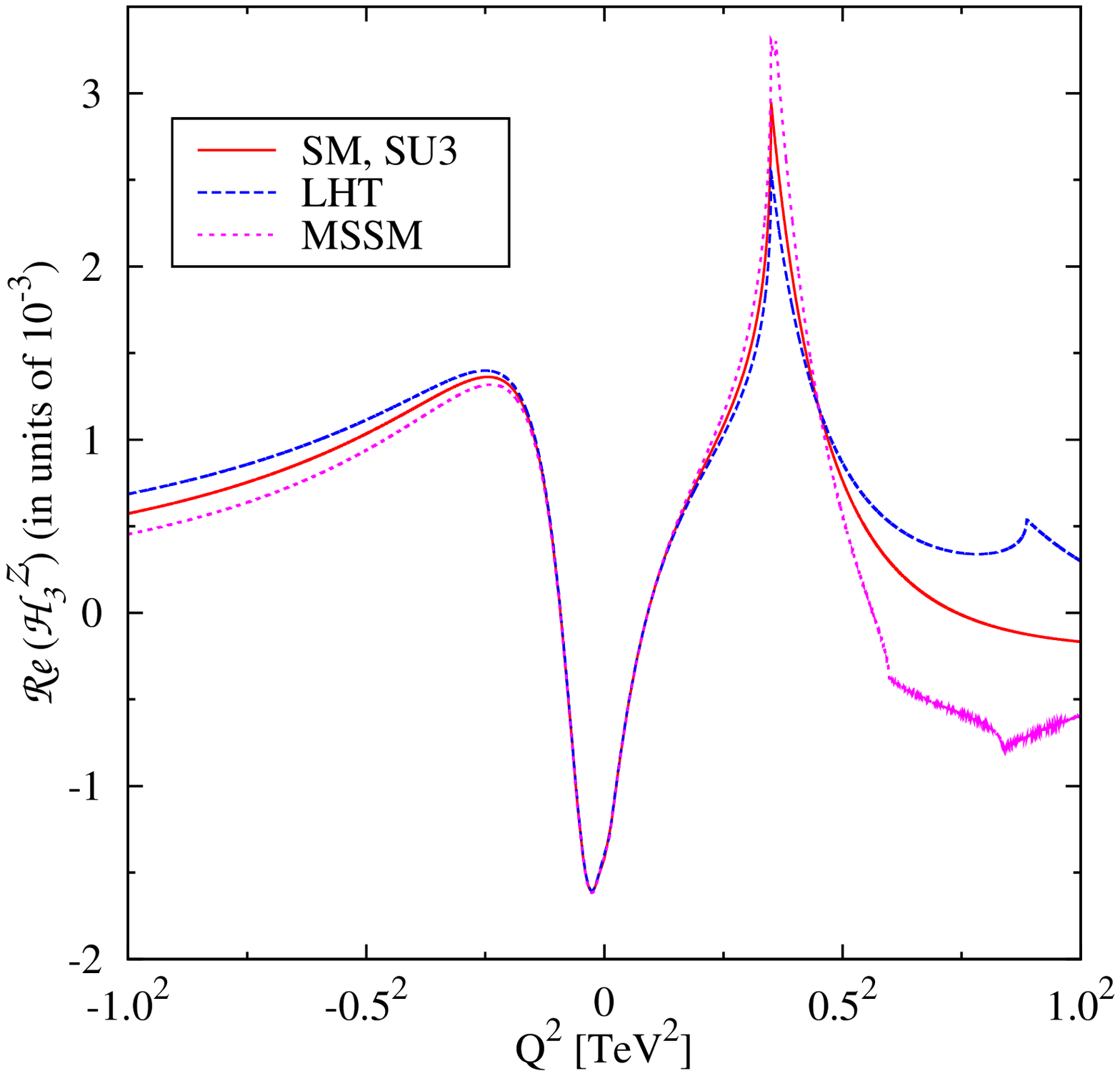}
   \end{minipage}
\vskip -4 cm 
\hskip -3 cm
  \begin{minipage}[t]{0.33\textwidth}
   \includegraphics[width=8cm,height=12cm]
   {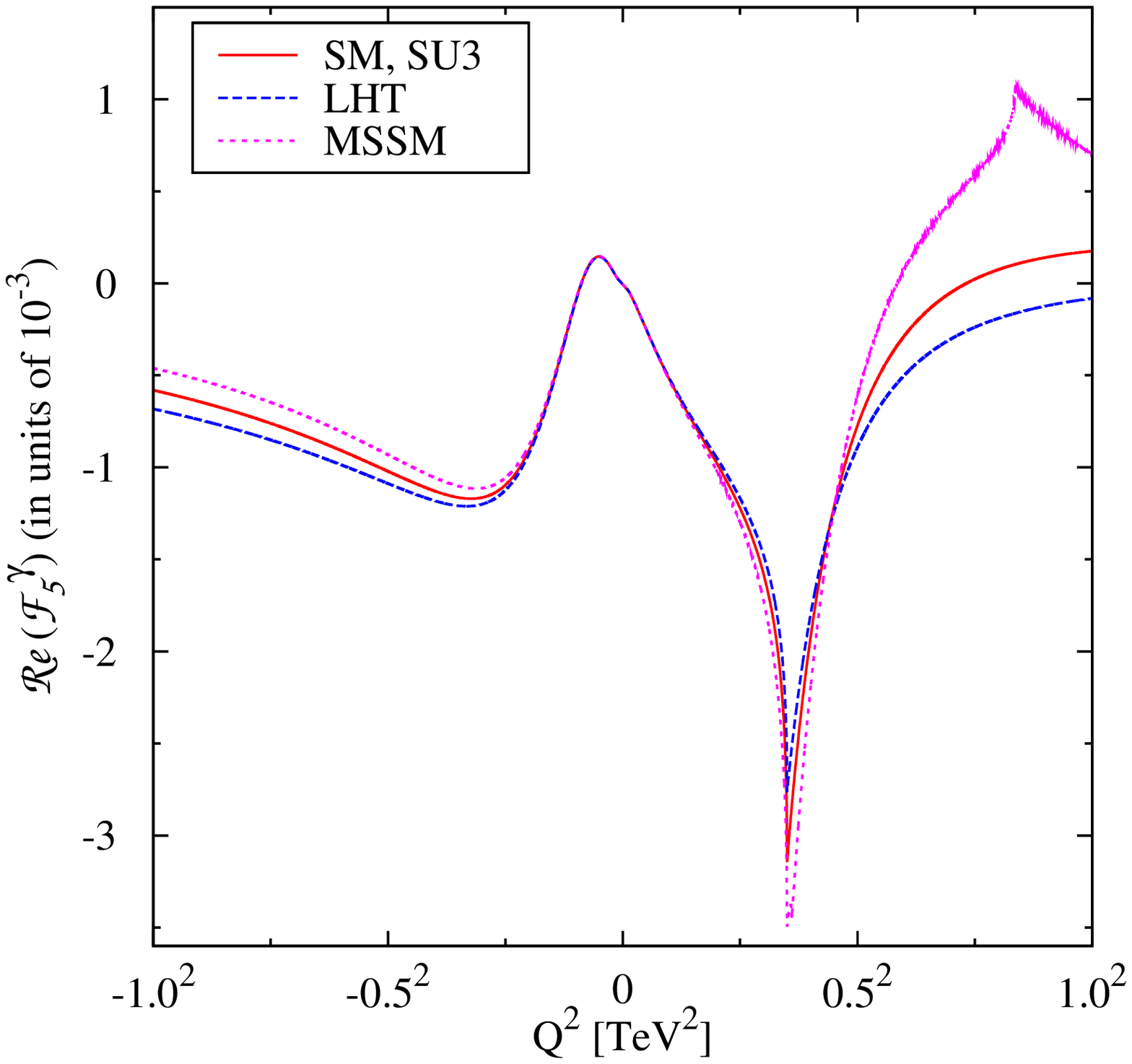}
  \end{minipage}
\hskip 3  cm
  \begin{minipage}[t]{0.33\textwidth}
   \includegraphics[width=8cm,height=12 cm]
   {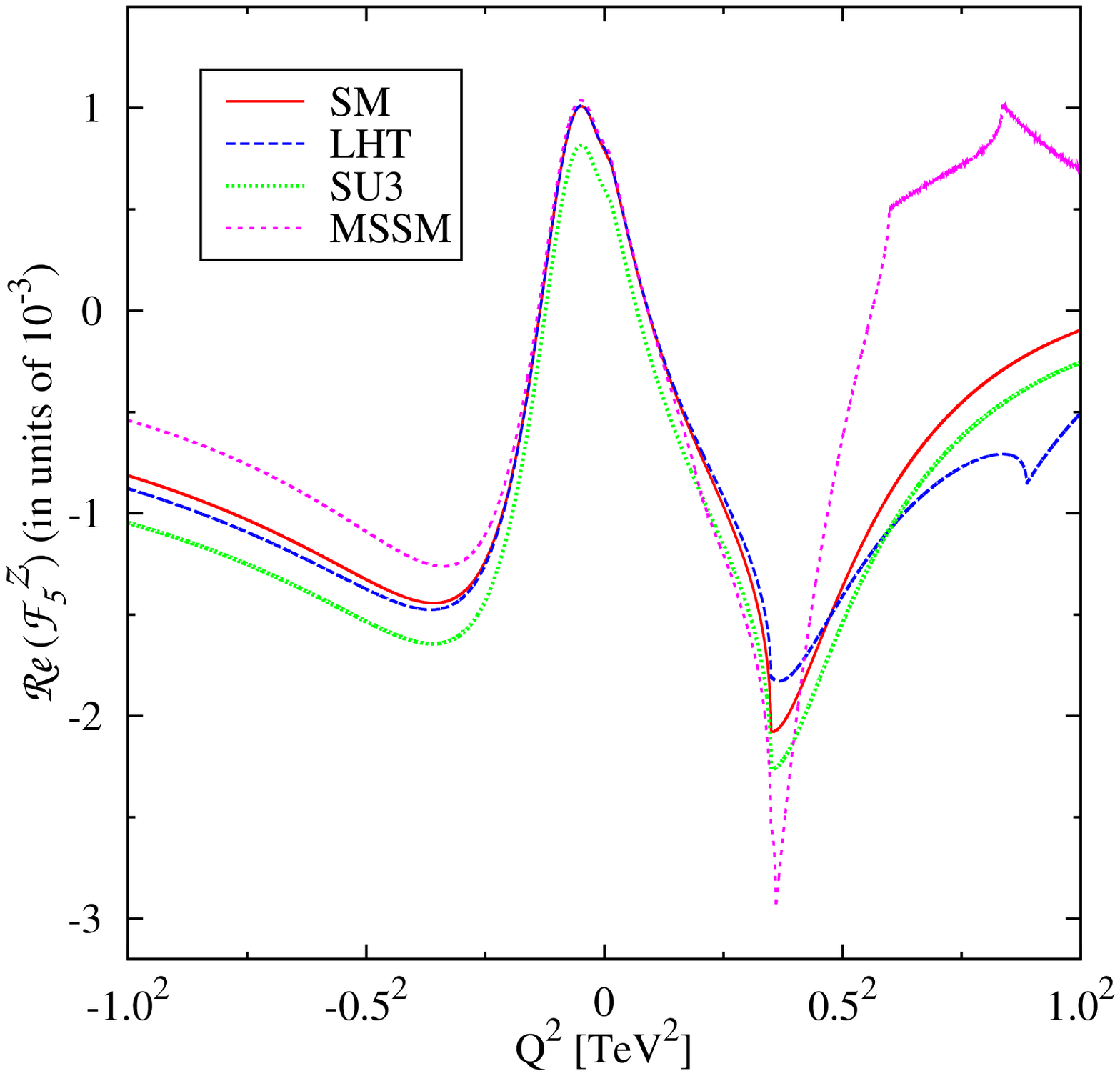}
  \end{minipage}
  \vskip -2cm
\caption{\small\em $\sqrt{Q^2}$-variation of the real parts of the couplings in
  various models for model parameter values $f=500$~GeV, $r=1$ for LHT and
  $f=3$~TeV, $t_\beta = 3$ and masses of all heavy fermions, $M_i = 2$~TeV for
  the SU(3) Model. MSSM parameters are as discussed in the text.}
  \label{CombinedRealParts}
\end{center}
\end{figure}
\begin{figure}[tbh!]
 \begin{center}
\hskip -3 cm
  \begin{minipage}[t]{0.33\textwidth}
   \includegraphics[width=8 cm,height=12 cm,]
   {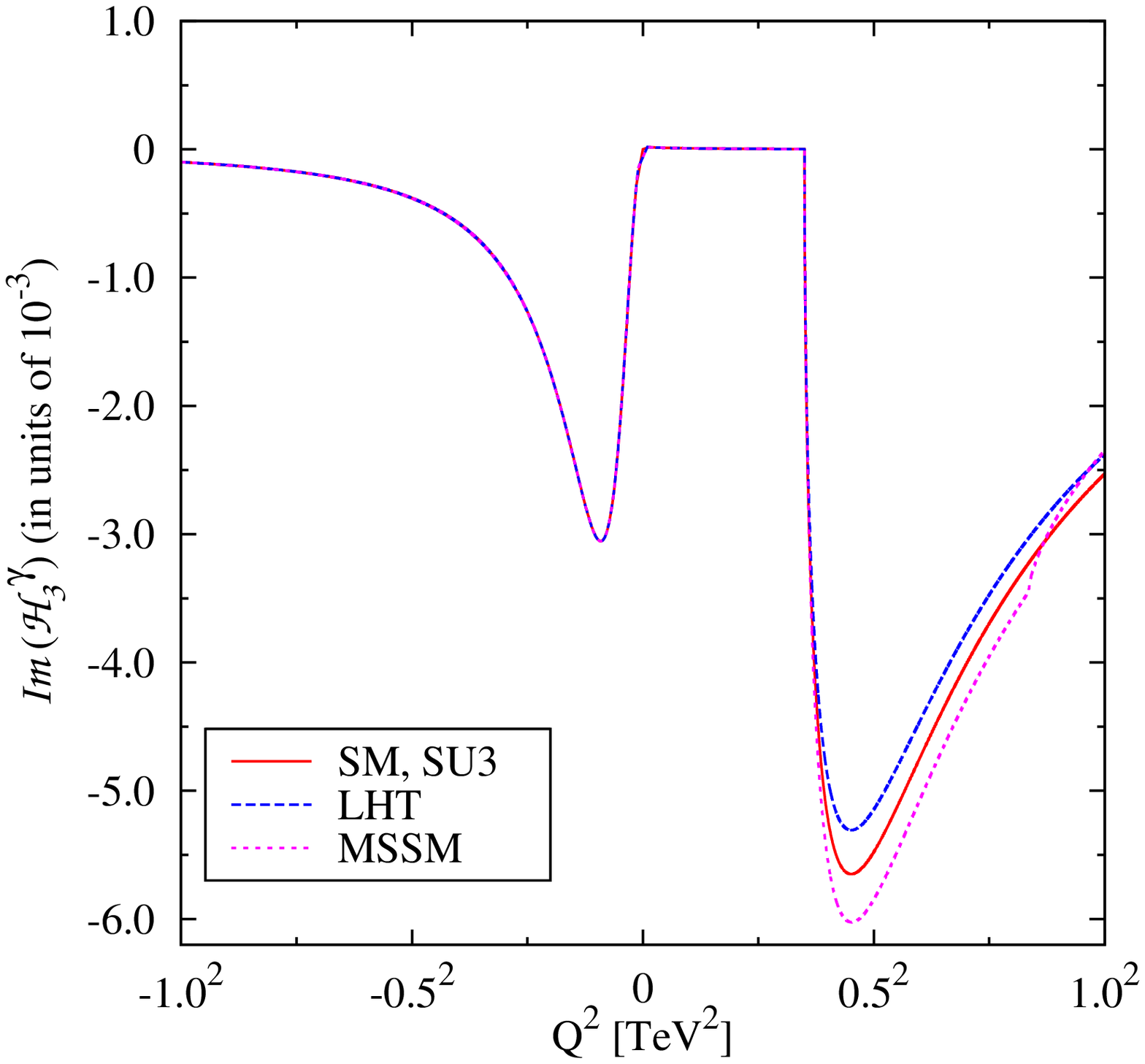}
  \end{minipage}
\hskip 3 cm
  \begin{minipage}[t]{0.33\textwidth}
   \includegraphics[width=8cm,height=12 cm]
   {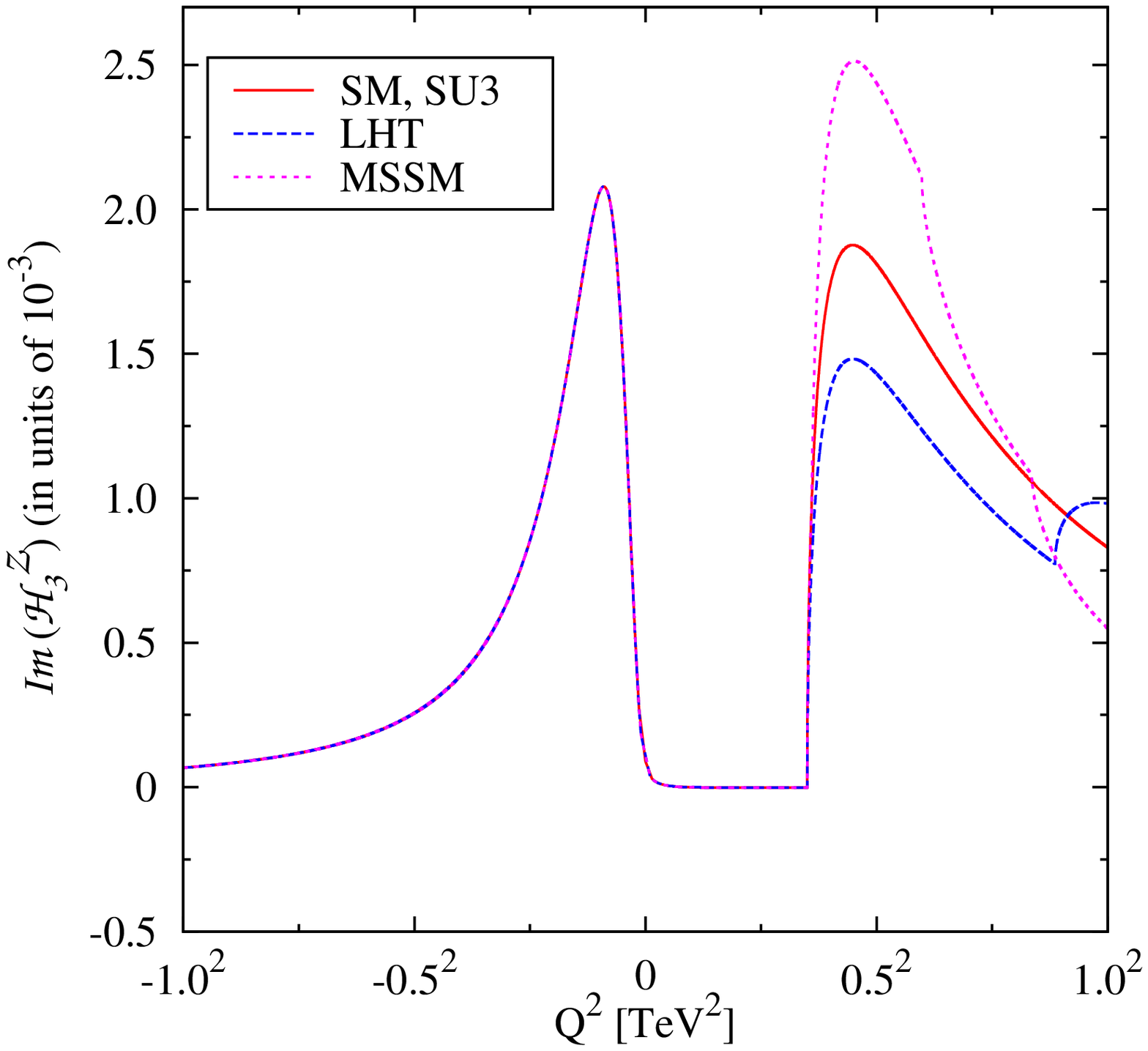}
   \end{minipage}
  \vskip -4 cm
\hskip -3 cm
  \begin{minipage}[t]{0.33\textwidth}
   \includegraphics[width=8cm,height=12cm]
   {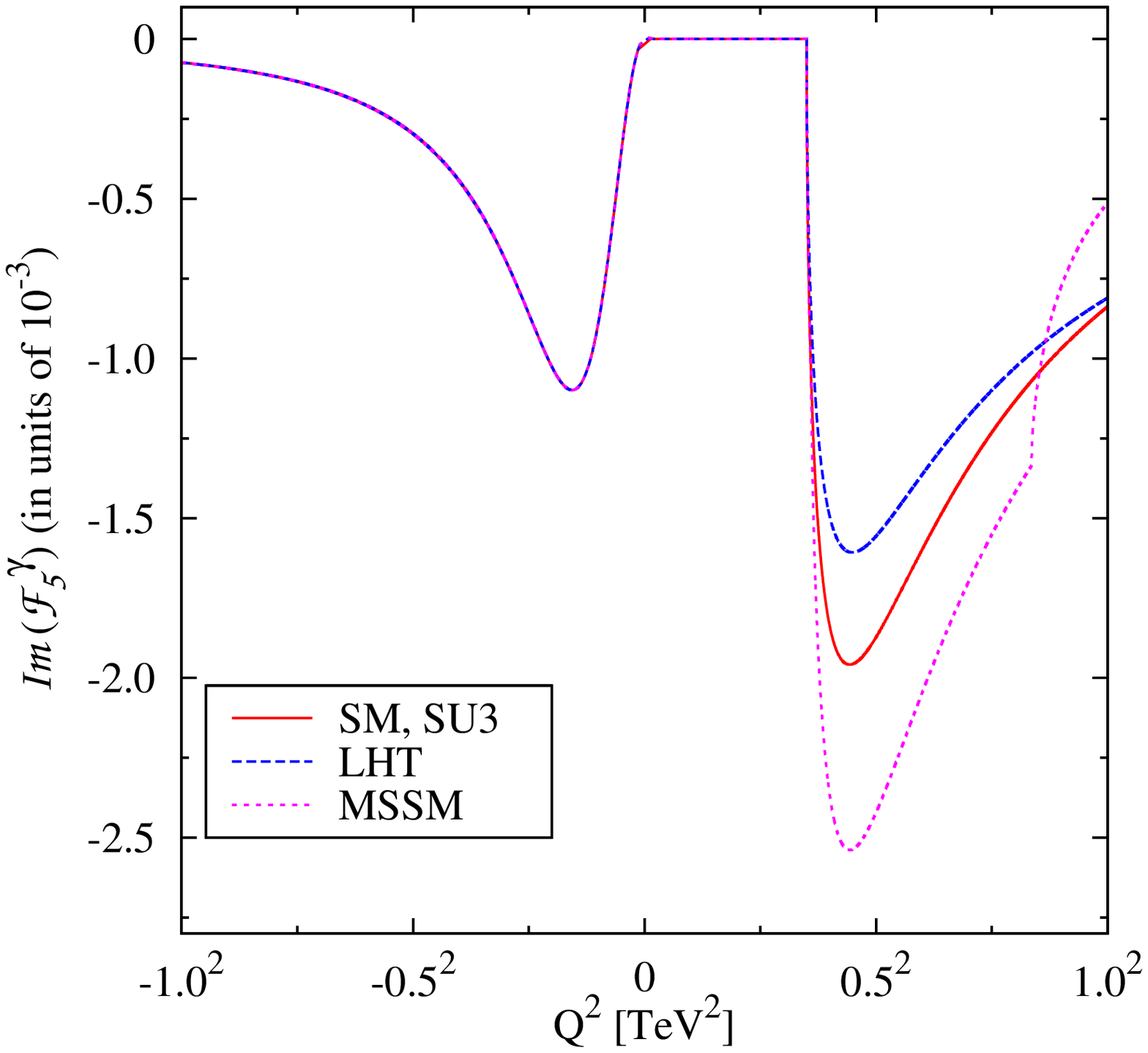}
  \end{minipage}
\hskip 3 cm
  \begin{minipage}[t]{0.33\textwidth}
   \includegraphics[width=8cm,height=12 cm]
   {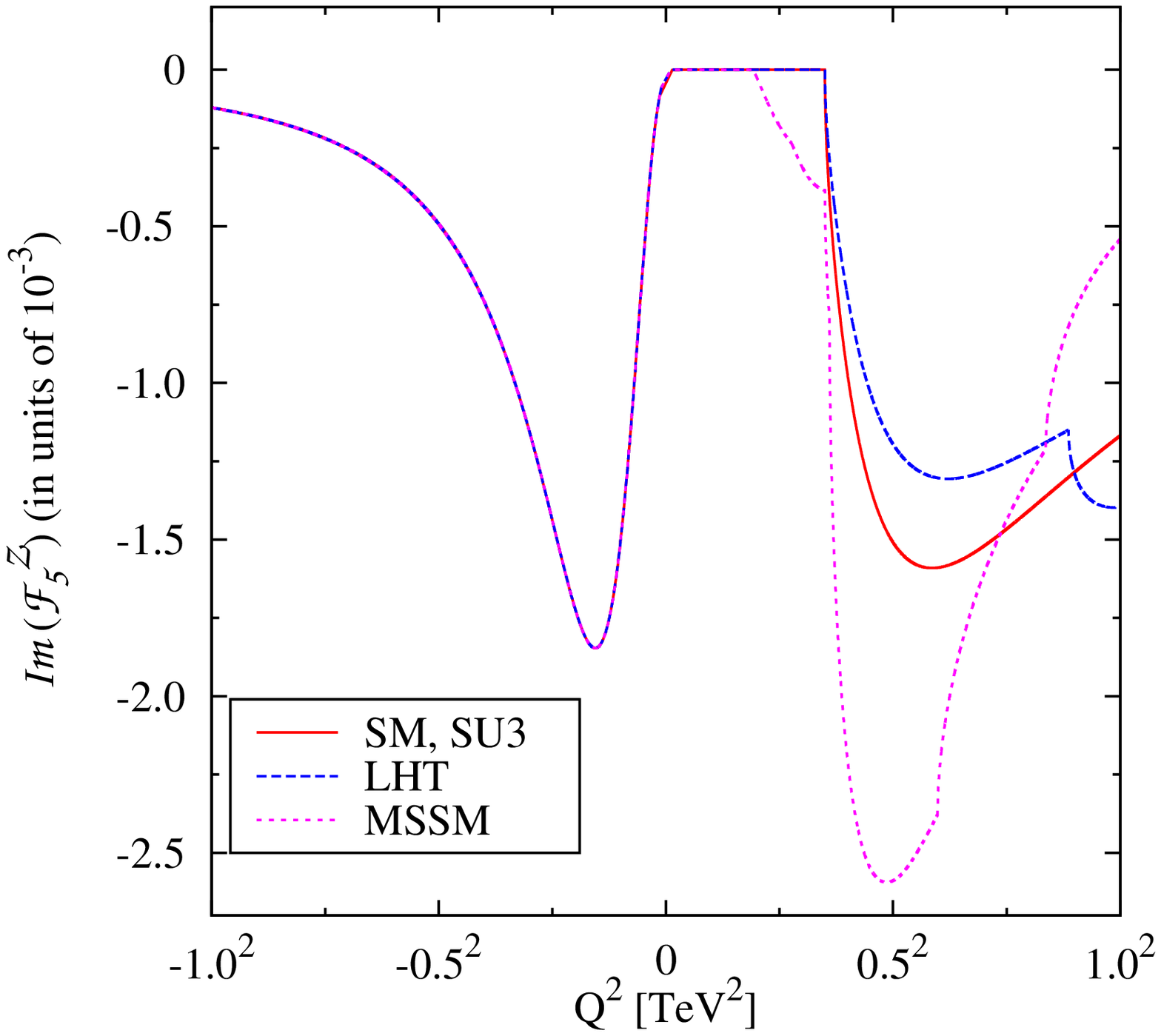}
  \end{minipage}
  \vskip -2cm
  \caption{\small\em $\sqrt{Q^2}$-variation of the imaginary parts of the couplings in
  various models for model parameter values $f=500$~GeV, $r=1$ for LHT and
  $f=3$~TeV, $t_\beta = 3$ and masses of all heavy fermions, $M_i = 2$~TeV for
  the SU(3) Model. MSSM parameters are as discussed in the text.}
  \label{CombinedImagParts}
 \end{center}
\end{figure}

We calculate the one-loop contribution to the CP-conserving trilinear
neutral gauge boson couplings in SM, MSSM and the two classes of
Little Higgs Models for various parameters of the models. The
$\sqrt{Q^2}$ variation of the real and imaginary parts of the
couplings in all the four models is shown in
Fig.~\ref{CombinedRealParts} and \ref{CombinedImagParts}. Values at
some typical $\sqrt{Q^2}$ are also given in the Tables~
\ref{tab:lht_scale_var}, \ref{tab:su3} and \ref{tab:lht_rat_var} for
different values of parameters of the model for the Little Higgs Model
with T-parity and for the SU(3) Model with anomaly free
embedding. Certain features are common to all these graphs which we
note here. All couplings vanish asymptotically for large $\sqrt{Q^2}$
compared to the highest fermion mass in the theory. This is ensured by
the anomaly cancellation in the models considered. The relative
importance of the real and imaginary parts of the couplings is
strongly energy dependent. As expected and explained in
Sec.~\ref{ngbc}, below the $2m_t$ threshold, the imaginary parts of all the
couplings are negligible. At and above this threshold the imaginary
parts become comparable or even dominant in comparison to the real
parts. This behavior is shared by the couplings in all the models
considered.

As discussed in Sec.~\ref{intro}, the triple gauge boson couplings in
the SM and MSSM have been already studied by various
groups~\cite{Choudhury:2000bw,Gounaris:1999kf}. Our results agree with
the earlier results. However we have chosen parameter space defined by
the reference point SPS1a$^\prime$ as mentioned in
Sec.~\ref{intro}. For this chosen point in the parameter space, the
chargino masses are calculated to be $m_{\chi_1^+} = 183.7 \gev$,
$m_{\chi_2^+} = 415.4 \gev$ and the neutralino masses are
$m_{\chi_1^0} = 94.8 \gev$, $m_{\chi_2^0} = 180.3 \gev$, $m_{\chi_3^0}
= 401.9 \gev$ and $m_{\chi_4^0} = 411.8 \gev$. In the MSSM there is a
peak at $\sqrt{Q^2}=2\, m_{\chi^+_1}$, which is very near to the $2\,
m_t$ SM peak for the special point chosen here. This results in the
enhancement of the couplings at this point as can be seen from
Figs.~\ref{CombinedRealParts} and  \ref{CombinedImagParts}. This
effect is more pronounced in the imaginary parts of the couplings. All
couplings in MSSM show a threshold effect at $\sqrt{Q^2}= 2\,
m_{\chi^+_2}\simeq $ 800 GeV which is more pronounced in the real
parts of $f_5^\gamma $ and $h_3^Z$ whereas in SM and Little Higgs
Model there is no such effect upto 1 TeV.
 Besides, in the MSSM  new peaks appear
at $m_{\chi_1^+} +m_{\chi_2^+}$ and $2 m_{\chi_2^+}$. As mentioned
before the neutralinos contribute only to $\f5z$.

The effect of extra heavy fermions in the LHT Model is to decrease the
threshold effects of the the SM whereas the particles in MSSM enhance
it. The new threshold in the LHT at $\sqrt{Q^2}=m_t+M_{T_+}$ and in the
MSSM as mentioned above are opposite to each other but the magnitudes
are comparable. As expected, the anomaly free SU(3) simple Model does
not show any appreciably different behavior than the SM upto $\sqrt{Q^2} =
1 \tev$.
\par We have studied the trilinear neutral gauge boson couplings
$\gamma^\star\,\gamma$, $Z^\star\,Z\,\gamma$, $\gamma^\star\, Z\,Z$
and $Z^\star\, Z\, Z$ involving one off-shell vector boson as a
function of $Q^2$ in SM, MSSM and Little Higgs models. We have made
theoretical prediction of these couplings for the model parameters
which are constrained by the electro-weak precision measurements. 
The large s-channel contributions in the $Z\,Z$ and $Z\,\gamma$ production at
the  LHC due to the anomalous triple gauge boson couplings, 
could be the first indirect manifestation of new physics. The $Z\,Z$ cross-section will be measured at the start up of LHC with a significance of 4.8  at the 1 fb$^{-1}$ integrated luminosity and expected to  suffer only a total of 12.9\%
uncertainties  which include the PDF and QCD uncertainties \cite{Ball:2007zza}. However, a  the precise measurement of the 
 triple gauge boson couplings will only be possible with a  10 fb$^{-1}$ luminosity. 
Our analysis presented above allows us to confront and discriminate among 
various models considered here on the basis of these couplings.
\par The complementary study of the   one-loop contribution to the triple charge - neutral gauge boson vertices  $W^+\,W^-\,\gamma$ and $W^+\,W^-\,Z$ in the context of   various Little Higgs Models with and without T parity  is in process.


\section*{Acknowledgments}
Authors would like to thank Prof. S. Rai Choudhury and Prof. Debajyoti Choudhury for fruitful discussions.
The authors acknowledge the partial support from the Department of Science
and Technology, India under grant SR/S2/HEP-12/2006  and the
infra-structural support from the IUCAA Reference Center, Delhi.
\appendix
\setcounter{equation}{0}
\section{Fermion 1 loop contribution to CP even Couplings}
        \label{sec:generic} 
\begin{figure}[tbh!]
 \begin{center}
\vskip -3 cm
\hskip -13 cm
  \begin{minipage}[t]{0.33\textwidth}
   \includegraphics[width=18 cm,height=21 cm]
   {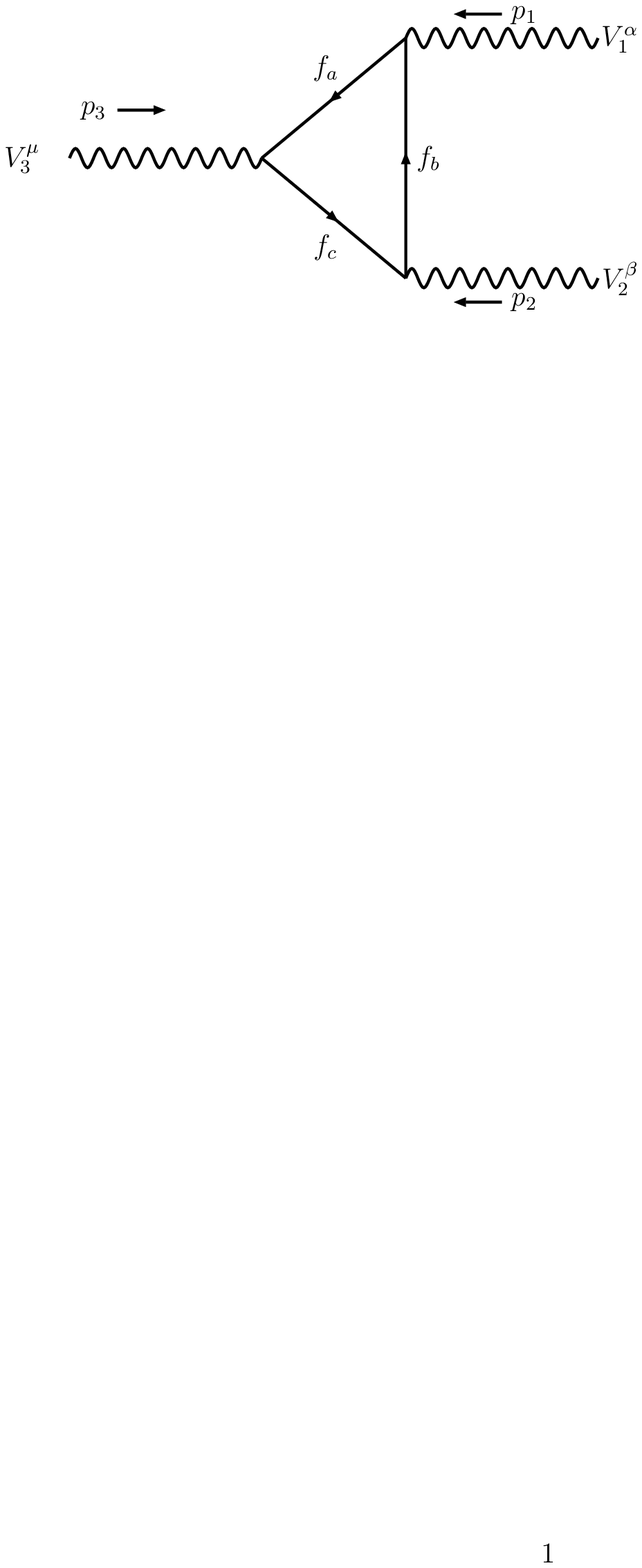}
  \end{minipage}
\hskip -3 cm
  \begin{minipage}[t]{0.33\textwidth}
   \includegraphics[width=18 cm,height=21 cm]
   {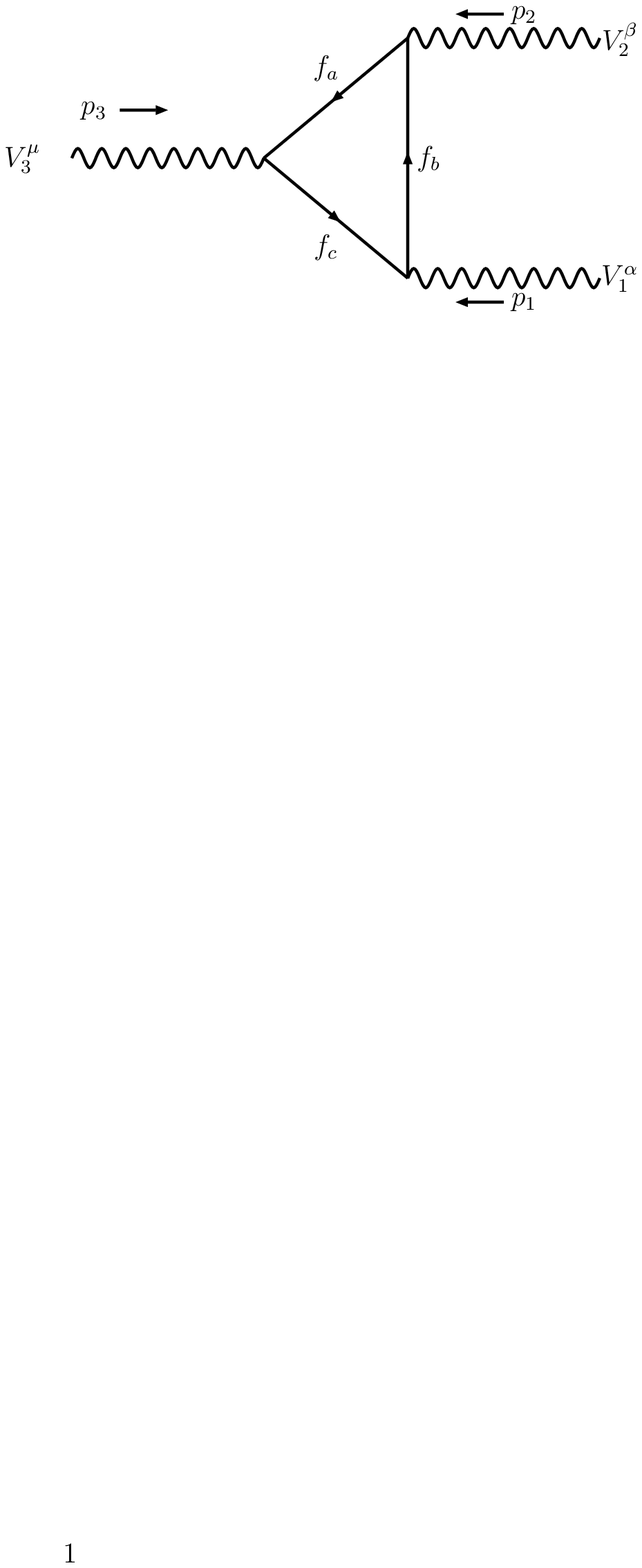}
   \end{minipage}

\end{center}
\vskip -15 cm
 (a) Direct \hskip 8 cm  (b) Exchange
\caption[]{\em Generic 1 loop diagrams contributing to CP conserving trilinear neutral gauge
boson vertices.}\label{fig:feyn}
\end{figure}

Let us consider the vertex: $V_1(p_{1\alpha})\, V_2(p_{2\beta})\,
V_3(p_{3\mu})$ where $V_i \equiv \gamma, Z$ and $p_1+p_2+p_3 = 0$. As for the CP conserving ones, to one-loop order, only the
fermions in the theory contribute. In
Fig.~\ref{fig:feyn}, we draw a generic direct and a exchange diagram
contributing to this process. Denoting the fermion-gauge coupling by
\begin{equation}
\mathcal L =\bar f_i \,\, \gamma_{\mu}\,\,\Big[{g_L}_{ij}^{\rm V} \,\,P_L + {g_R}_{ij}^{\rm V}\,\,P_R\Big] f_j\,\, {\rm V}^{\mu}\, .
\end{equation}

with $P_{L, R} = (1 \mp \gamma_5) / 2$. Throughout our analysis the momenta $p_1$ and $p_2$ denote
same kind of bosons, say ${\rm V}_1^\alpha$ and ${\rm V}_2^\beta$, and momenta $p_3$  denotes
${\rm V}_3^\mu$ which is of the second kind of vector boson unlike the case of $ZZZ$, where all are  vector boson are same.
 It is useful to define the
combinations   in notation where ${\rm V}\equiv {\rm V}_1\equiv {\rm V}_2$.
\be \barr{rcl} g_{1}^{{\rm V},\,{\rm V}_3} & = &\,\,\,\,\,\,\,\,\,\,\,\,\,\,\,\,\,\,\,\, \dis  \left( {g_L}_{ab}^{\rm V} \,\,\, {g_L}_{bc}^{\rm V}  \,\,\,
{g_L}_{ca}^{{\rm V}_3} 
 -  {g_R}_{ab}^{\rm V}  \,\,\, {g_R}_{bc}^{\rm V}  \,\,\, {g_R}_{ca}^{{\rm V}_3} \right), \\
 g_2 ^{{\rm V},\,{\rm V}_3}& = & \dis m_a\,
m_c \,\,\,\left({g_L}_{ab}^{\rm V} \,\,\, {g_L}_{bc}^{\rm V}  \,\,\,
{g_R}_{ca}^{{\rm V}_3}   - {g_R}_{ab}^{\rm V} \,\,\, {g_R}_{bc}^{\rm V}  \,\,\,
{g_L}_{ca}^{{\rm V}_3} \right), \\ 
g_3^{{\rm V},\,{\rm V_3}} & = & \dis m_a \, m_b \,\,\, \left( {g_L}_{ab}^{\rm V} \,\,\, {g_R}_{bc}^{\rm V}  \,\,\,
{g_R}_{ca}^{{\rm V}_3} - {g_R}_{ab}^{\rm V} \,\,\, {g_L}_{bc}^{\rm V}  \,\,\,
{g_L}_{ca}^{{\rm V}_3}\right), \\
g_4 ^{{\rm V},\,{\rm V}_3}& = &
\dis m_b \, m_c\,\,\, \left( {g_L}_{ab}^{\rm V} \,\,\, {g_R}_{bc}^{\rm V}  \,\,\,
{g_L}_{ca}^{{\rm V}_3} - {g_R}_{ab}^{\rm V} \,\,\, {g_L}_{bc}^{\rm V}  \,\,\,
{g_R}_{ca}^{{\rm V}_3} \right).   \earr \label{the_combins}   
\ee
Here $m_i$, $i=a,b,c$, are the masses of the internal fermions $f_i$.
The loop contribution with three distinct internal fermions $f_a$,
$f_b$, $f_c$ having masses $m_a$, $m_b$ and $m_c$, respectively,
corresponding to the direct diagram of Fig.~\ref{fig:feyn}(a) is
proportional to $\epsilon_{\alpha \beta \mu \eta} $ and can be
parameterized as
\renewcommand{\thefootnote}{\fnsymbol{footnote}}$^{\footnotemark[1]}$
\footnotetext[1]{We adopt the convention $\epsilon_{0123}=1$. Also we
 follow the notation of PV functions as given in Ref.~\cite{pv}}
\be
\barr{rcl}
{\cal A}^{\cal D}_{a,b,c} ({\rm V},\,{\rm V}_3;\, \, p_1,\,p_2,\,p_3)&
   = & \dis g_1^{{\rm V},\,{\rm V}_3} \Big[ p_3^2 (C_{11}+C_{21}) -
   p_2^{2}(C_{12} + C_{22}) - 2 p_1\cdot p_2 (C_{12}+C_{23}) \Big]
   \\[1.5ex] &+ & \dis (m_a^2 g_1^{{\rm V},\,{\rm V}_3} + g_3^{{\rm
   V},\,{\rm V}_3} - g_2^{{\rm V},\,{\rm V}_3}) (C_{11}+C_{0}) -
   g_4^{{\rm V},\,{\rm V}_3} C_{11}.  \earr
\label{eq:loop_d}
\ee
The contribution of exchange diagram with three distinct fermions
$f_a$, $f_b$, $f_c$ as shown in the Fig.~\ref{fig:feyn}(b) having
 masses $m_a$, $m_b$ and $m_c$, respectively and proportional
to the $\epsilon_{\alpha \beta \mu \eta}$ is parameterized in terms of
\be
\barr{rcl}
{\cal A}^{\cal E}_{a,b,c} ({\rm V},\,{\rm V}_3; \,\,p_1,\,p_2,\,p_3) & = & \dis
   - (m_a^2 g_1^{{\rm V},\,{\rm V}_3} - g_4^{{\rm V},\,{\rm V}_3} -
g_2^{{\rm V},\,{\rm V}_3}) (C_{11}-C_{12}) 
             - g_3^{{\rm V},\,{\rm V}_3} (C_{11}+C_{0}-C_{12})
  \\[1.5ex]  
          & & - \,\dis 
g_1^{{\rm V},\,{\rm V}_3} \Big[B_{023}+ B_{123} -2 C_{24} 
                + p_1^{2} (C_{12}+C_{21} )
  \\[1.5ex]  
        &  & \dis  \hspace*{4em}
                + (p_2^{2} + 2 p_1 \cdot p_2)
(C_{22}-2 C_{23}+C_{21}) \Big]
 \\[2ex]
\earr\label{eq:loop_ex}
 \ee
In Eqs.~\ref{eq:loop_d} and \ref{eq:loop_ex}, $B_{\mu jk}$ and
$C_\mu, \, C_{\mu\,\nu}$ are respectively the 2-point and 3-point
Passarino Veltman functions (\cite{pv}) defined as
\bea
B_{023} = B_0(p^2_2,\, m^2_b,\, m^2_c);\,\,\, B_{123} = B_1(p^2_2,\, m^2_b,\, m^2_c);\nonumber\\
C_0; \, C_\mu;\, C_{\mu \nu} = C_0; \, C_\mu;\, 
C_{\mu \nu}(p^2_3,\, p^2_2,\, m^2_a,\, \, m^2_c,\, m^2_b) 
\label{b-fn_not} 
\eea
with
\bea
B_0;B_\mu; B_{\mu \nu}(p^2, m^2_i,m^2_j) &=& \frac{1}{\iota\, \pi^2} \int{d^4 k \frac{1;k_\mu;k_{\mu \nu}}{(k^2 +m_i^2)[(k+p)^2 + m_j^2]}}  \nonumber \\
C_0;C_\mu; C_{\mu \nu}(p^2_1,p^2_2, m^2_i,m^2_j,m^2_k) &=& \frac{1}{\iota\, \pi^2} \int{d^4 k \frac{1;k_\mu;k_{\mu \nu}}{(k^2 +m_i^2)[(k+p_1)^2 + m_j^2][(k+p_1+p_2)^2 + m_k^2]}} \nonumber \\
\label{bc-fn_def}
\eea
\begin{enumerate}
\item The loop amplitude for $\gamma^\star \gamma Z$  is
\be
\kappa\,\,\sum_{a=1}^{N_f} \sum_{b=1}^{N_f}\sum_{c=1}^{N_f} {\cal C}_f\,\,\bigg[{{\cal A}^{\cal D}_{a,b,c} }(\gamma,\,Z; \,\,0,\,\sqrt{Q^2},\,m_Z) +{{\cal A}^{\cal E}_{a,b,c} }(\gamma,\,Z; \,\sqrt{Q^2},\,0,\,m_Z)\bigg],
\ee
\noindent where ${N_f}$ is the total number of flavors, ${\cal C}_f$ is the color factor of  the fermion in the loop and $\kappa$ is the over all loop factor.
\noindent Since electromagnetic interactions at two  vertices  forbids  any flavor mixing, the above summation  can be re-written as
\begin{eqnarray}
{\cal H}_3^\gamma&=&-\,\left(\frac{4\,\pi\alpha_{\rm em}}{\big(2\,\cos\theta_W\,\sin\theta_W\big)\big(16\,\pi^2\big)}\right)\,\,\otimes\nonumber\\
&&\sum_{a=1}^{N_f}  {\cal C}_f\,\,\bigg[{{\cal A}^{\cal D}_{a,a,a} }(\gamma,\,Z; \,\,0,\,\sqrt{Q^2},\,m_Z) +{{\cal A}^{\cal E}_{a,a,a} }(\gamma,\,Z; \,\sqrt{Q^2},\,0,\,m_Z)\bigg].
\end{eqnarray}

\item The loop amplitude for $\gamma^\star Z Z$  has one e.m. vertex which renders the mixing among the weak interaction eigenstates at  the other two weak   vertices giving
\begin{eqnarray}
{\cal F}_5^\gamma&=&\left(\frac{4\,\pi\alpha_{\rm em}}{\big(2\,\cos\theta_W\,\sin\theta_W\big)\big(16\,\pi^2\big)}\right)\,\otimes\nonumber\\
&&\,\sum_{a=1}^{N_f} \sum_{b=1}^{N_f}  {\cal C}_f\,\,\bigg[{{\cal A}^{\cal D}_{a,b,a} }(Z,\,\gamma; \,\,m_Z,\,m_Z,\,\sqrt{Q^2}) +
{{\cal A}^{\cal E}_{a,b,a} }(Z,\,\gamma; \,\,m_Z,\,m_Z,\,\sqrt{Q^2}) \bigg]
\end{eqnarray}
\item The loop amplitude for $Z^\star \gamma Z$  follows the same suit as the previous one   with an additional change  in the momentum assignment.
\begin{eqnarray}
{\cal H}_3^Z&=&\left(\frac{4\,\pi\alpha_{\rm em}}{\big(2\,\cos\theta_W\,\sin\theta_W\big)^2\big(16\,\pi^2\big)}\right)\,\otimes\nonumber\\
&&\,\sum_{a=1}^{N_f} \sum_{b=1}^{N_f}  {\cal C}_f\,\,\bigg[{{\cal A}^{\cal D}_{a,b,a} }(Z,\,\gamma; \,\,m_Z,\, \sqrt{Q^2},\,0) 
+ {{\cal A}^{\cal E}_{a,b,a} }(Z,\,\gamma; \, \sqrt{Q^2}, \,m_Z,\,0)\bigg]
\end{eqnarray}
\item The loop amplitude for $Z^\star Z Z$  allow weak mixing at all vertices.
\begin{eqnarray}
{\cal F}_5^Z&=&\left(\frac{4\,\pi\alpha_{\rm em}}{\big(2\,\sin \theta_W\,\cos\theta_W \big)^3\,\big(16\,\pi^2\big)}\right)\,\otimes\nonumber\\
&&\sum_{a=1}^{N_f} \sum_{b=1}^{N_f}\sum_{c=1}^{N_f} {\cal C}_f\,\, \bigg[{{\cal A}^{\cal D}_{a,b,c} }(Z,\,Z; \,\,m_Z,\, m_Z,\,\sqrt{Q^2})
+{{\cal A}^{\cal E}_{a,b,c} }(Z,\,Z; \,\,m_Z,\, m_Z,\,\sqrt{Q^2}) \bigg]\nonumber\\
\end{eqnarray}
\end{enumerate}
\begin{table}[!h]
\begin{center}
\begin{tabular}{||c||c|c||c|c||}
\hline
\hline
Vertex & \multicolumn{2}{c|}{LHT Model} & \multicolumn{2}{||c||}{SU(3)
simple group} \\
\cline{2-5}
& $g_L$ & $g_R$  &$g_L$ & $g_R$ \\
\hline
\hline
$\bar q_i Z q_i $ &&&&  \\
(for $i=1-5$) & $2(T_3^i - Q^i s_w^2)$ & $-2 Q^i s_w^2$ & 
$2(T_3^i - Q^i s_w^2)$ & $-2 Q^i s_w^2$ \\
 i.e. all SM quarks except top &&&& \\
\hline
$\bar l_i Z l_i $ &&&&  \\
(for $i=1-6$) & $2(T_3^i - Q^i s_w^2)$ & $-2 Q^i s_w^2$ & 
$2(T_3^i - Q^i s_w^2)$ & $-2 Q^i s_w^2$ \\
 i.e. all SM leptons &&&& \\
\hline
$\bar t Z t $ & $1 - \frac{4}{3} s_w^2 - x_L^2 \frac{v^2}{f^2} $   
 & $- \frac{4}{3} s_w^2 $ & 
 $1 - \frac{4}{3} s_w^2 - x_L^2 \frac{v^2}{f^2} $   
 & $- \frac{4}{3} s_w^2 $ \\
\hline
$\bar T Z t $ & $ x_L \frac{v}{f} $   
 & 0 & 
$\frac{1}{2 \sqrt{2} (r^2 +t_\beta^2 )} s_{2 \beta} (1 + t_\beta^2)(r^2 -1)\frac{v}{f} $
 & $0 $ \\
\hline
$\bar T Z T $ & $ x_L^2 \frac{v^2}{f^2} - \frac{4}{3}  s_w^2   $   
 & $ - \frac{4}{3}  s_w^2   $ &  $0 $ & $0 $ \\
\hline
$\bar D_i Z d_i $ &&&& \\
$ i= 1,2$ & $ \times \times $   
 & $  \times \times  $ &  $ \frac{1}{ \sqrt{2} t_\beta} \frac{v}{f} $ & $0 $ \\
\hline
$\bar N_i Z \nu_i $ &&&& \\
$ i= 1,2,3 $ & $ \times \times $   
 & $  \times \times  $ &  $- \frac{1}{ \sqrt{2} t_\beta} \frac{v}{f} $ & $0 $ \\\hline
\hline
\end{tabular}
\caption{ \label{tab:coup} {\em Relevant Couplings of fermions with
Z-boson in units of $ g/2c_w$ in the Littlest Higgs Model with
T-parity (LHT) and SU(3) simple group model with anomaly-free
embedding. Note that T denotes the T-even heavy top quark in case of
LHT and is the heavy top in case of SU(3) Model.  Here $x_L =
\frac{1}{1+ r^{-2}}$ with $ r = \lambda_1/\lambda_2$.}}
\end{center}
\end{table}




\begin{thebibliography}{99}
\bibitem{Ball:2007zza}
  G.~L.~Bayatian {\it et al.}  [CMS Collaboration],
  J.\ Phys.\ G {\bf 34}, 995 (2007).
\bibitem{Djouadi:2007ik}
  A.~Djouadi {\it et al.}  [ILC Collaboration],
  arXiv:0709.1893 [hep-ph].


\bibitem{Gounaris:1996rz}
  G.~Gounaris {\it et al.},
  arXiv:hep-ph/9601233;
  E.~N.~Argyres, A.~B.~Lahanas, C.~G.~Papadopoulos and V.~C.~Spanos,
  Phys.\ Lett.\  B {\bf 383}, 63 (1996)
  [arXiv:hep-ph/9603362];
  E.~N.~Argyres, G.~Katsilieris, A.~B.~Lahanas, C.~G.~Papadopoulos and V.~C.~Spanos,
  Nucl.\ Phys.\  B {\bf 391}, 23 (1993).
  A.~Arhrib, J.~L.~Kneur and G.~Moultaka,
  Phys.\ Lett.\  B {\bf 376}, 127 (1996)
  [arXiv:hep-ph/9512437].

\bibitem{Gounaris:1999kf}
  G.~J.~Gounaris, J.~Layssac and F.~M.~Renard,
  Phys.\ Rev.\  D {\bf 61}, 073013 (2000)
  [arXiv:hep-ph/9910395].
         
\bibitem{Choudhury:2000bw}
  D.~Choudhury, S.~Dutta, S.~Rakshit and S.~Rindani,
  Int.\ J.\ Mod.\ Phys.\  A {\bf 16}, 4891 (2001)
  [arXiv:hep-ph/0011205].

\bibitem{guzzi}  R. Armillis, C. Coriano and M. Guzzi, JHEP {\bf 0805}, 15 (2008) [arXiv:0711.3424 [hep-ph]].
\bibitem{baur}  U. Baur and E.L. Berger, Phys. Rev. {\bf D47}, 4889 (1993); U. Baur, T. Han and J. Ohnemus, Phys. Rev. {\bf D57}, 2823 (1998); U. Baur and D. Rainwater, Phys. Rev. {\bf D52}, 112011 (2000) and references therein.

\bibitem{Deng:2008zz} J.~Deng, FERMILAB-THESIS-2008-37, (2008); 
  T.~Aaltonen {\it et al.}  [CDF Collaboration],
  Phys.\ Rev.\  D {\bf 76}, 111103 (2007)
  [arXiv:0705.2247 [hep-ex]].
\bibitem{:2007pq}
  J.~Abdallah {\it et al.}  [DELPHI Collaboration],
  Eur.\ Phys.\ J.\  C {\bf 51}, 525 (2007)
  [arXiv:0706.2741 [hep-ex]];
  M.~Acciarri {\it et al.}  [L3 Collaboration],
  Phys.\ Lett.\  B {\bf 489} (2000) 55
  [arXiv:hep-ex/0005024].


\bibitem{ArkaniHamed:2002qx}
  N.~Arkani-Hamed, A.~G.~Cohen, E.~Katz, A.~E.~Nelson, T.~Gregoire and J.~G.~Wacker,
  JHEP {\bf 0208}, 021 (2002)
  [arXiv:hep-ph/0206020].

 
\bibitem{ArkaniHamed:2002qy}
  N.~Arkani-Hamed, A.~G.~Cohen, E.~Katz and A.~E.~Nelson,
  JHEP {\bf 0207}, 034 (2002)
  [arXiv:hep-ph/0206021].

\bibitem{Low:2002ws}
  I.~Low, W.~Skiba and D.~Tucker-Smith,
  Phys.\ Rev.\  D {\bf 66}, 072001 (2002)
  [arXiv:hep-ph/0207243];
  S.~Chang and J.~G.~Wacker,
  Phys.\ Rev.\  D {\bf 69}, 035002 (2004)
  [arXiv:hep-ph/0303001];

  S.~Chang,
  JHEP {\bf 0312}, 057 (2003)
  [arXiv:hep-ph/0306034].

\bibitem{Skiba:2003yf}
  W.~Skiba and J.~Terning,
  Phys.\ Rev.\  D {\bf 68}, 075001 (2003)
  [arXiv:hep-ph/0305302].

         
\bibitem{Kaplan:2003uc}
  D.~E.~Kaplan and M.~Schmaltz,
  JHEP {\bf 0310}, 039 (2003)
  [arXiv:hep-ph/0302049].

\bibitem{Schmaltz:2004de}
  M.~Schmaltz,
  JHEP {\bf 0408}, 056 (2004)
  [arXiv:hep-ph/0407143].

\bibitem{Csaki:2003si}
  C.~Csaki, J.~Hubisz, G.~D.~Kribs, P.~Meade and J.~Terning,
  Phys.\ Rev.\  D {\bf 68}, 035009 (2003)
  [arXiv:hep-ph/0303236];
  J.~L.~Hewett, F.~J.~Petriello and T.~G.~Rizzo,
  JHEP {\bf 0310}, 062 (2003)
  [arXiv:hep-ph/0211218].

         \bibitem{Barbieri:2004qk}
  R.~Barbieri, A.~Pomarol, R.~Rattazzi and A.~Strumia,
  Nucl.\ Phys.\  B {\bf 703}, 127 (2004)
  [arXiv:hep-ph/0405040]; Z.~Han and W.~Skiba,
  Phys.\ Rev.\  D {\bf 72}, 035005 (2005)
  [arXiv:hep-ph/0506206].
\bibitem{marandella}
  G.~Marandella, C.~Schappacher and A.~Strumia,
  Phys.\ Rev.\  D {\bf 72}, 035014 (2005)
  [arXiv:hep-ph/0502096].

\bibitem{Cheng:2003ju}
  H.~C.~Cheng and I.~Low,
  JHEP {\bf 0309}, 051 (2003)
  [arXiv:hep-ph/0308199].
    

\bibitem{Hubisz:2004ft}
  J.~Hubisz and P.~Meade,
  Phys.\ Rev.\  D {\bf 71}, 035016 (2005)
  [arXiv:hep-ph/0411264];
  J.~Hubisz, S.~J.~Lee and G.~Paz,
  JHEP {\bf 0606}, 041 (2006)
  [arXiv:hep-ph/0512169];
  J.~Hubisz, P.~Meade, A.~Noble and M.~Perelstein,
  JHEP {\bf 0601}, 135 (2006)
  [arXiv:hep-ph/0506042].
 

\bibitem{Choudhury:2004bh}
  S.~R.~Choudhury, N.~Gaur, A.~Goyal and N.~Mahajan,
  Phys.\ Lett.\  B {\bf 601}, 164 (2004)
  [arXiv:hep-ph/0407050];
  M.~Blanke, A.~J.~Buras, A.~Poschenrieder, C.~Tarantino, S.~Uhlig and A.~Weiler,
  JHEP {\bf 0612}, 003 (2006)
  [arXiv:hep-ph/0605214];
  M.~Blanke, A.~J.~Buras, A.~Poschenrieder, S.~Recksiegel, C.~Tarantino, S.~Uhlig and A.~Weiler,
  JHEP {\bf 0701}, 066 (2007)
  [arXiv:hep-ph/0610298];
  A.~Goyal,
  Mod.\ Phys.\ Lett.\  A {\bf 21}, 1931 (2006);
  S.~R.~Choudhury, N.~Gaur and A.~Goyal,
  Phys.\ Rev.\  D {\bf 72}, 097702 (2005)
  [arXiv:hep-ph/0508146];
  M.~Blanke, A.~J.~Buras, B.~Duling, A.~Poschenrieder and C.~Tarantino,
  JHEP {\bf 0705}, 013 (2007)
  [arXiv:hep-ph/0702136].
\bibitem{hagiwara} K. Hagiwara, R. D. Pecci, D. Zepanfeld and K. Hikasa, Nucl. Phys. {\bf B282}, 253 (1987).
\bibitem{Han:2005ru}
  T.~Han, H.~E.~Logan and L.~T.~Wang,
  JHEP {\bf 0601}, 099 (2006)
  [arXiv:hep-ph/0506313].

 \bibitem{simplemod1}
R.~Barbieri, A.~Pomaral, R.~Rattazi and A.~Strumia, Nucl. Phys. {\bf B 703},
127 (2004).
\bibitem{spa} J. A. Aguilar-Saavedra {\it et. al.}, EPJC (2006) 
 arXiv:hep-ph/0511344V1 
\bibitem{pv}
G.~Passarino and M.~Veltman,  Nucl. Phys. {\bf B 160}, 151 (1971).
\end{thebibliography}
\end{document}